\documentclass[pdflatex,sn-nature]{sn-jnl}%

\usepackage{CJKutf8}
\usepackage{graphicx}%
\usepackage{multirow}%
\usepackage{amsmath,amssymb,amsfonts}%
\usepackage{aas_macros}
\usepackage{amsthm}%
\usepackage{mathrsfs}%
\usepackage[title]{appendix}%
\usepackage[dvipsnames]{xcolor}
\usepackage{textcomp}%
\usepackage{manyfoot}%
\usepackage{booktabs}%
\usepackage{algorithm}%
\usepackage{algorithmicx}%
\usepackage{algpseudocode}%
\usepackage{listings}%
\usepackage{textgreek}
\usepackage{xargs}
\usepackage{xspace}

\newcommand{\Halpha}{\text{H\textalpha}\xspace}
\newcommand{\Hbeta}{\text{H\textbeta}\xspace}

\newcommandx{\permittedEL}[6][1=O,2=III,3=,4=,5=,6=]{\text{{#1}\,{\sc{#2}}{#3}{#4}{#5}{#6}}\xspace}
\newcommandx{\semiforbiddenEL}[6][1=O,2=III,3=,4=,5=,6=]{\text{{#1}\,{\sc{#2}}]{#3}{#4}{#5}{#6}}\xspace}
\newcommandx{\forbiddenEL}[6][1=O,2=III,3=,4=,5=,6=]{\text{[{#1}\,{\sc{#2}}]{#3}{#4}{#5}{#6}}\xspace}

\newcommandx{\HII}{\permittedEL[H][ii]}
\newcommandx{\HeI}{\permittedEL[He][i]}
\newcommandx{\HeIL}[1][1=3889]{\permittedEL[He][i][\,\textlambda][#1]}
\newcommandx{\HeIIL}[1][1=4686]{\permittedEL[He][ii][\,\textlambda][#1]}
\newcommand{\OIII}{\forbiddenEL[O][iii]}
\newcommandx{\OIIIL}[1][1=5007]{\forbiddenEL[O][iii][\textlambda][#1]}

\newcommandx{\OIIIlow}[1][1=1666]{\semiforbiddenEL[O][iii][\textlambda][#1]}
\newcommandx{\NIL}[1]{\forbiddenEL[N][i][\textlambda][5200]}

\newcommandx{\OIIL}[1][1=3727]{\forbiddenEL[O][ii][\textlambda][#1]}

\newcommand{\NII}{\forbiddenEL[N][ii]}
\newcommandx{\NIIL}[1][1=6583]{\forbiddenEL[N][ii][\textlambda][#1]}

\newcommandx{\CIVall}{\permittedEL[C][iv][\textlambda][\textlambda][1549,][1551]}
\newcommandx{\CIV}{\permittedEL[C][iv]}

\newcommandx{\NV}{\permittedEL[N][v]}
\newcommandx{\NeV}{\forbiddenEL[Ne][v]}
\newcommand{\arcsec}{$^{\prime\prime}$}

\newcommand{\edit}[1]{#1}

\theoremstyle{thmstyleone}%
\theoremstyle{thmstyletwo}%

\theoremstyle{thmstylethree}%

\begin{document}

\title[An Ancient Descendant of The First Galaxies]{An Ancient Descendant of The First Galaxies}

\author*[1, 2, 8]{\fnm{Jacqueline} \sur{Antwi-Danso}}\email{j.antwidanso@utoronto.ca}
\author[3]{\fnm{Adam} \sur{Muzzin}}
\author[4]{\fnm{Luke} \sur{Robbins}}
\author[5,6,8]{\fnm{Yoshihisa} \sur{Asada}}
\author[4]{\fnm{Danilo} \sur{Marchesini}}
\author[6]{\fnm{Marcin} \sur{Sawicki}}
\author[7]{\fnm{Kartheik} \sur{Iyer}}
\author[2]{\fnm{Katherine} \sur{Whitaker}}
\author[1,8, 9, 10]{\fnm{Joshua S.} \sur{Speagle (\begin{CJK*}{UTF8}{bsmi}沈佳士\ignorespacesafterend\end{CJK*})}}
\author[11, 12]{\fnm{Casey} \sur{Papovich}}
\author[13]{\fnm{Chris} \sur{Willott}}
\author[14]{\fnm{Maru\v{s}a} \sur{Brada\v{c}}}
\author[15]{\fnm{Guillaume} \sur{Desprez}}
\author[14]{\fnm{Vladan} \sur{Markov}}
\author[14]{\fnm{Nicholas} \sur{Martis}}
\author[16]{\fnm{Ga\"{e}l} \sur{Noirot}}
\author[3]{\fnm{Ghassan T.E.} \sur{Sarrouh}}
\author[3]{\fnm{Rahul} \sur{Kannan}}
\author[1]{\fnm{Roberto} \sur{Abraham}}
\author[1]{\fnm{Seiji} \sur{Fujimoto}}
\author[3]{\fnm{Katherine} \sur{Myers}}

\affil*[1]{David A. Dunlap Department of Astronomy and Astrophysics, University of Toronto, 50 St. George Street, Toronto, Ontario, M5S 3H4, Canada}

\affil[2]{\orgname{Department of Astronomy, University of Massachusetts, Amherst}, \orgaddress{710 N Pleasant Street}, \city{Amherst, MA}, \postcode{01003}, \country{USA}}

\affil[3]{\orgdiv{Department of Physics and Astronomy}, \orgname{York University}, \orgaddress{\street{4700 Keele St.}, \city{Toronto}, \postcode{M3J 1P3}, \state{Ontario}, \country{Canada}}}

\affil[4]{\orgdiv{Department of Physics \& Astronomy}, \orgname{Tufts University}, \orgaddress{\street{574 Boston Avenue}, \city{Medford, MA}, \postcode{02155}, \country{USA}}}

\affil[5]{\orgdiv{Department of Astronomy}, \orgname{Kyoto University}, \orgaddress{\street{Sakyo-ku}, \city{Kyoto}, \postcode{606-8502}, \state{Kyoto}, \country{Japan}}}

\affil[6]{\orgdiv{Department of Astronomy and Physics and Institute for Computational Astrophysics}, \orgname{Saint Mary's University}, \orgaddress{\street{923 Robie Street}, \city{Halifax}, \postcode{B3H 3C3}, \state{Nova Scotia}, \country{Canada}}}

\affil[7]{\orgdiv{Columbia Astrophysics Laboratory}, \orgname{Columbia University}, \orgaddress{\street{550 West 120th Street}, \city{New York}, \postcode{10027}, \state{NY}, \country{USA}}}

\affil[8]{\orgname{Dunlap Institute for Astronomy and Astrophysics}, \orgaddress{50 St. George Street}, \city{Toronto, Ontario}, \postcode{M5S 3H4}, \country{Canada}}

\affil[9]{\orgname{Department of Statistical Sciences, University of Toronto}, \orgaddress{700 University Ave}, \city{Toronto, ON}, \postcode{M5G 1Z5}, \country{Canada}}

\affil[10]{\orgname{Data Sciences Institute, University of Toronto}, \orgaddress{700 University Ave}, \city{Toronto, ON}, \postcode{M5G 1Z5}, \country{Canada}}

\affil[11]{\orgname{George P. and Cynthia Woods Mitchell Institute for Fundamental Physics and Astronomy, Texas A\&M University}, \orgaddress{576 University Dr}, \city{College Station, TX}, \postcode{78743}, \country{USA}}

\affil[12]{\orgname{Department of Physics and Astronomy, Texas A\&M University}, \orgaddress{4242 TAMU}, \city{College Station, TX}, \postcode{78743}, \country{USA}}

\affil[13]{NRC Herzberg, 5071 West Saanich Rd, Victoria, BC V9E 2E7, Canada}

\affil[14]{\orgdiv{Faculty of Mathematics and Physics}, \orgname{University of Ljubljana}, \orgaddress{\street{19 Jadranska ulica}, \city{Ljubljana}, \postcode{1000}, \country{Slovenia}}}

\affil[15]{\orgdiv{Kapteyn Astronomical Institute, University of Groningen},  \orgaddress{\street{P.O. Box 800}, \city{Groningen}, \postcode{9700AV}, \country{The Netherlands}}}

\affil[16]{\orgdiv{Space Telescope Science Institute}, \orgaddress{\street{3700 San Martin Drive}, \city{Baltimore, Maryland}, \postcode{21218}, \country{USA}}}



\maketitle

\textbf{JWST has revealed unexpectedly bright galaxies in the first 500 Myr \edit{after the Big Bang \cite{donnan2024, chmerynska2024}}. \edit{Their overabundance suggests that they are preferentially observed during burst phases \cite{sun2023}, where their star formation rates increase dramatically \cite{cole2025}. In cosmological simulations, such bursts transition into short ($\lesssim 40$ Myr) periods without star formation or  ``naps" \cite{dome2024, gelli2025}. }Using JWST/NIRCam medium-band observations, we report the discovery of the galaxy CANUCS-A370-2228423 ($z=5.95 \pm 0.06$, log($M_\ast/M_\odot) = 9.14 \pm 0.09$), dubbed ``\emph{The Sleeper}." Its star formation history \edit{indicates rapid assembly in the first 300 Myr ($z \gtrsim 14$)}, where it formed a log($M_\ast/M_\odot) = 8.7^{+0.3}_{-0.4}$ $M_\odot$ progenitor, comparable in stellar mass to the few spectroscopically-confirmed galaxies at those redshifts \cite{carniani2024, carniani2025, robertson2023, naidu2025, castellano2024}. \edit{Unexpectedly, this is followed by several hundred million years of suppressed star formation, in stark contrast to ``nappers" \cite{looser2024, witten2025, vikaeus2024, strait2023}. This results in a} remarkably strong hydrogen Balmer break, exceeding that of any galaxy observed within the first billion years \edit{by a factor of $\approx 3$}. \edit{Furthermore, Sleeper-like systems are overabundant in the observed survey volume compared to theory, as the probability of finding such galaxies in simulations is $< 0.2\%$.} \edit{The discovery of \emph{The Sleeper} therefore disrupts the current narrative that all luminous galaxies in the first few hundred million years grow into massive descendants. Instead it presents an alternative evolutionary pathway in which these unusually luminous galaxies fade into inefficient dwarfs after an early starburst, revealing greater diversity in the first stages of galaxy evolution.}
}


\edit{Cosmological simulations predict that luminous galaxies at cosmic dawn grow into the most massive galaxies in the present-day Universe \cite{lu2025}. }Given their  stellar masses ($M_\ast \approx 10^8 - 10^9 ~\mathrm{M_\odot}$) and high star-formation rates \cite{harikane2023, robertson2023, naidu2025, castellano2024}, it has been postulated that these UV-bright galaxies at $ z > 10$ are progenitors of \edit{objects like} RUBIES-UDS-QG-z7 \cite{weibel2025}, a \edit{recently-discovered} massive (log($M_\ast/M_\odot) \approx 10.2$) quiescent galaxy at $z \approx 7.3$. \edit{Reproducing the inferred star formation history of this system requires gas-rich mergers at $z \approx 8 - 9$ that trigger intense starbursts of $\sim 100~ \mathrm{M_\odot ~ yr^{-1}}$, a plausible scenario since these galaxies} occupy the highest-density peaks of the early Universe  \cite{paquereau2025}. This predicts extremely UV-bright ($\mathrm{M_{UV}} \lesssim -24$) systems at these redshifts, consistent with abundance matching arguments, which suggest a luminosity evolution of $\mathrm{M_{UV}} \propto (1 + z)^{-4.5}$ \cite{carniani2024}. Despite their predicted high number densities, no such UV-bright descendants have been observed at $z > 4$ to date. \edit{Analytical models \cite{wyithe2014} instead predict a second evolutionary pathway whereby luminous galaxies fade rapidly after short-lived bursts, leaving behind UV-faint relics at $z \approx 4 - 8$ with low-to-intermediate stellar masses ($10^9 - 10^{9.5}~\mathrm{M_\odot}$) and suppressed star formation for several 100 Myr. These extended lulls are in contrast to the phases of temporary quiescence or ``mini-quenching" prevalent in the low-mass population \cite{gelli2025, looser2024, dome2024, witten2025}.} Consequently, these relics are dominated by relatively red stellar populations (with rest-frame UV spectral slopes, $-2 \lesssim \beta \lesssim -1$). Their predicted duty cycle of $10-15\%$ further implies they are more abundant at fixed stellar mass than currently observed \cite{wyithe2014}.  

\begin{figure}[t]%
\centering
\includegraphics[trim={0.1cm 0.65cm 0cm 0.1cm},clip,width=\textwidth]{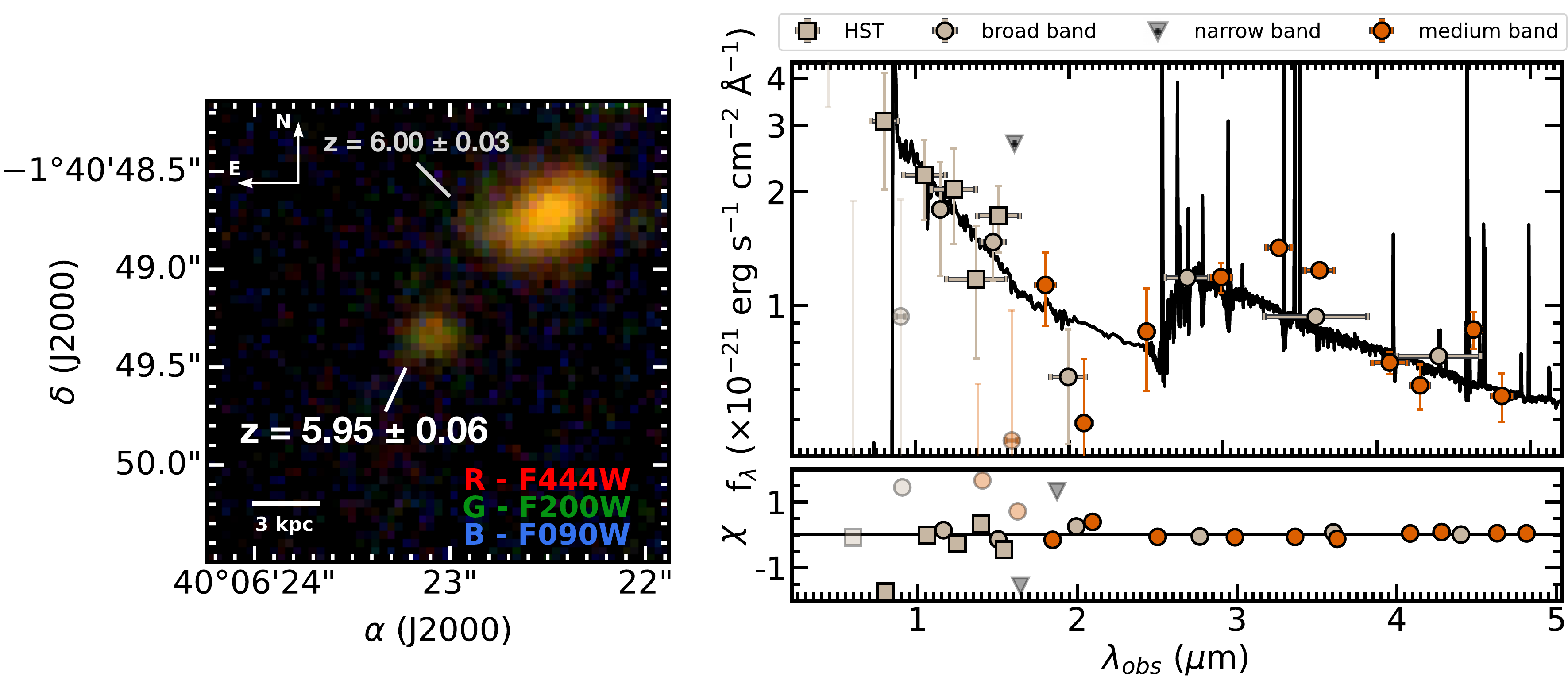}
\caption{Photometry and spectral energy distribution (SED) modeling of \emph{The Sleeper}. Left: \emph{JWST}/NIRCam color-composite image. \emph{The Sleeper} is located at a projected distance of $\approx 3$ kpc from a more massive ($\log(M_\star/M_\odot) = 9.55 \pm 0.35$) companion (CANUCS-A370-2201097), both situated in an overdensity at $z = 5.87 - 6.14$ with 36 \edit{other galaxies showing Balmer breaks}. Right: maximum likelihood {\tt DENSE BASIS} fit ($\chi_\nu^2 = 1.14$) to the 29-band photometry from CANUCS/Technicolor. This comprises all the available \emph{JWST}/NIRCam medium and broad bands, combined with two narrow bands and ancillary \emph{HST} data. The F250M and F300M medium bands, \edit{as well as the low fluxes in F200W and F210M}, robustly identify a strong flux excess caused by the redshifted hydrogen Balmer spectral break ($\lambda_{\mathrm{rest}} = 3645~$\AA), which unambiguously indicates an evolved population when the observed SED is dominated by stars. 
}\label{fig:sed}
\end{figure}

\edit{In this paper, we report observations of CANUCS-A370-2228423 (``\emph{The Sleeper}"). In line with predictions from analytical models, it is characterized by a faint absolute magnitude of M$_\mathrm{UV} = -17.32 \pm 0.44$ ($\approx 0.04L^{\star}$) \cite{willott2024} and a red UV slope ($\beta = -1.58 \pm 0.46$), making it the first reported candidate relic of UV-luminous galaxies at $z > 10$.} \emph{The Sleeper} was identified using imaging  from the Canadian NIRISS Unbiased Cluster Survey (CANUCS) \cite{willott2022}  and \emph{JWST in Technicolor} programs combined with archival \emph{Hubble Space Telescope (HST)} observations \cite{lotz2017} (see Methods). This dataset utilizes all broad- and medium-bands, and two narrow bands available on \emph{JWST}/NIRCam at $\sim 1 - 5~\mu$m. The galaxy's extensive 29-band photometric SED is shown in Figure \ref{fig:sed}. \emph{The Sleeper} was identified as having a strong Balmer break at $\lambda_{rest} = 3645$~\AA, detected as a large flux excess between the F250M and F300M bands. The low flux in the adjacent F200W and F210M bands relative to the aforementioned further corroborate the existence of a spectral break. Due to the detection of the $\lambda_{rest} = 1216$~\AA~ Lyman break, Balmer break, and rest-optical emission lines, the redshift is robustly constrained to $z = 5.95 \pm 0.06$ with the {\tt DENSE BASIS} \cite{iyer2017, iyer2019} code, \edit{in agreement with other codes to within $\Delta z = +0.02$ (see Methods)}. The best-fit model ($\chi_\nu^2 = 1.1$) is able to reproduce the strong Balmer break with a stellar population dominated by late-type B-stars and A-stars. The presence of a red, evolved stellar population is evident in the galaxy's false-color image in Figure \ref{fig:sed}.

We measured the strength of the Balmer break, $\mathrm{D_B} = F_{\nu, 4225}/ F_{\nu, 3565}$, from the best-fit spectral energy distribution (SED) using the flux ratio using two emission line-free windows at $\lambda_{\rm rest} = 3500 - 3630$~\AA~ and $\lambda_{\rm rest} = 4160 - 4290$~\AA~ \cite{rb2024}. In the first billion years, following a burst of star formation, $\mathrm{D_B}$ increases rapidly as young, massive stars die off (Robbins et al., in prep). It is strongest in $\sim 500$ Myr old stellar populations, corresponding to the main-sequence turn off for type A2 V stars \cite{gonzelez-delgado1999}. This makes the Balmer break both an age indicator and a proxy for the quenching of star formation. \emph{The Sleeper}'s extreme Balmer break, $\mathrm{D_B} = 2.9_{-0.2}^{+0.3}$, is $\approx 3\times$ higher than the typical measured at $z \sim 5-7$ \cite{endsley2024, rb2024}. Indeed, only $1\%$ of spectroscopically-confirmed galaxies at these redshifts have $\mathrm{D_B} > 1.4$ \cite{covelo-paz2025}. Even in the large volume \textsc{Sphinx$^{20}$} ($20^3$ cMpc$^3$) \cite{katz2023}, \textsc{Flares} (3.2 cGpc$^3$ parent volume) \cite{lovell2021}, and \textsc{thesan} ($95.5^3$ cMpc$^3$) \cite{kannan2022} hydrodynamical cosmological simulations, there is only a $<0.2\%$ probability of finding a galaxy with \emph{The Sleeper}'s break strength (Methods). Notably, it resides in a group of 37 galaxies at $z = 5.87 - 6.14$ (Methods), the majority of which (97\%) exhibit pronounced Balmer breaks, with a $1\sigma$ range of $\mathrm{D_B} \approx 1.1 - 1.9$. These strengths suggest stellar populations that are already several hundred million years old, indicating that the structure is unusually evolved for this epoch. This is earlier than expected from simulations \cite{chiang2017}, which find that galaxies in overdensities at $z \sim 6$ are still vigorously forming stars and hence dominated by young stellar populations. This is further corroborated by the \textsc{Flares} and \textsc{thesan} simulations, where we find that the typical Balmer break strength for $\approx 10^9~\mathrm{M_\odot}$ galaxies at $z = 6$ is $\mathrm{D_B \approx 1.2}$ (Methods). 


Although Balmer breaks can be mimicked by strong emission lines in high redshift galaxies when only photometry is available (e.g., \cite{desprez2024}), the medium bands from \emph{JWST} robustly distinguish a bonafide break from nebular emission by providing the equivalent of coarse spectroscopy, with a spectral resolution of $R \sim 15$ \cite{suess2024}. While the flux excess between F250M and F300M can be reproduced with a sufficiently strong emission line (W$_0 (\Hbeta) \gtrsim 2000$ \AA~\edit{at $z = 7.0$}), our models show that this is highly disfavored ($\chi^2_\nu = 4.3$), requiring a \edit{near-pristine ($Z = 0.016 Z \odot$)} stellar population to reproduce the observed SED. This is unlikely given that the required metallicity is comparable to those of the most metal-poor objects observed in the epoch of reionization \cite{willott2025, hsiao2025}. Moreover, highly star-forming models struggle to match the observed UV photometry, as extremely low metallicities are accompanied by weaker UV continuum due to stronger nebular emission. \edit{Therefore, strongly star-forming solutions (with specific star formation rates, log $\text{sSFR}_{10} [ ~\textrm{yr}^{-1}] \geq -7.5$) are excluded as poor fits at $99.9$\% credibility}. 

\edit{Dust-reddened models are also inconsistent with the data}. Although dust can boost the strength of the Balmer break, the best-fit dusty model requires $\approx 1$ mag of extinction. This yields up to $\sim 7 \sigma$ offsets in several bands covering the UV continuum, resulting in a worse fit ($\chi_\nu^2 = 2.3$) across the fitted data points compared to the fiducial ($\chi_\nu^2 = 1.1$). The data therefore strongly prefer low-to-moderate dust extinction ($\mathrm{A_V} \approx 0.2 - 0.4$ mag), irrespective of the adopted dust law or prior and stellar library (see Methods). These results are consistent with the finding that both observed \cite{algera2023} and simulated $\sim 10^9~ \mathrm{M_\odot}$ galaxies \cite{md2023, zimmerman2024} at $z \sim 4-6$ exhibit little dust obscuration. \edit{Given the low inferred extinction of \emph{The Sleeper}, we find that dust contributes only mildly to the measured Balmer break ($\Delta \mathrm{D_B} = +0.18$). Another way to reconcile the strong UV continuum with the prominent Balmer break is via a spatially-varying dust geometry \cite{faisst2024}. In this scenario, the observed SED is a composite consisting of UV flux from a star-forming region with little-to-no dust and a Balmer break arising from an obscured post-starburst. Cosmological simulations similarly attribute much of the scatter in dust extinction at fixed mass to spatial and viewing angle variations \cite{zimmerman2024}, though the observational evidence is limited e.g., \cite{bowler2024}. Having established that emission line contamination or significant dust obscuration is unlikely in \emph{The Sleeper}, we adopt the interpretation of an unambiguous Balmer break of remarkable strength and consider its implications.} 

\begin{figure}[t]%
\centering
\includegraphics[trim={0cm 1cm 0.cm 1cm}, width=\textwidth]{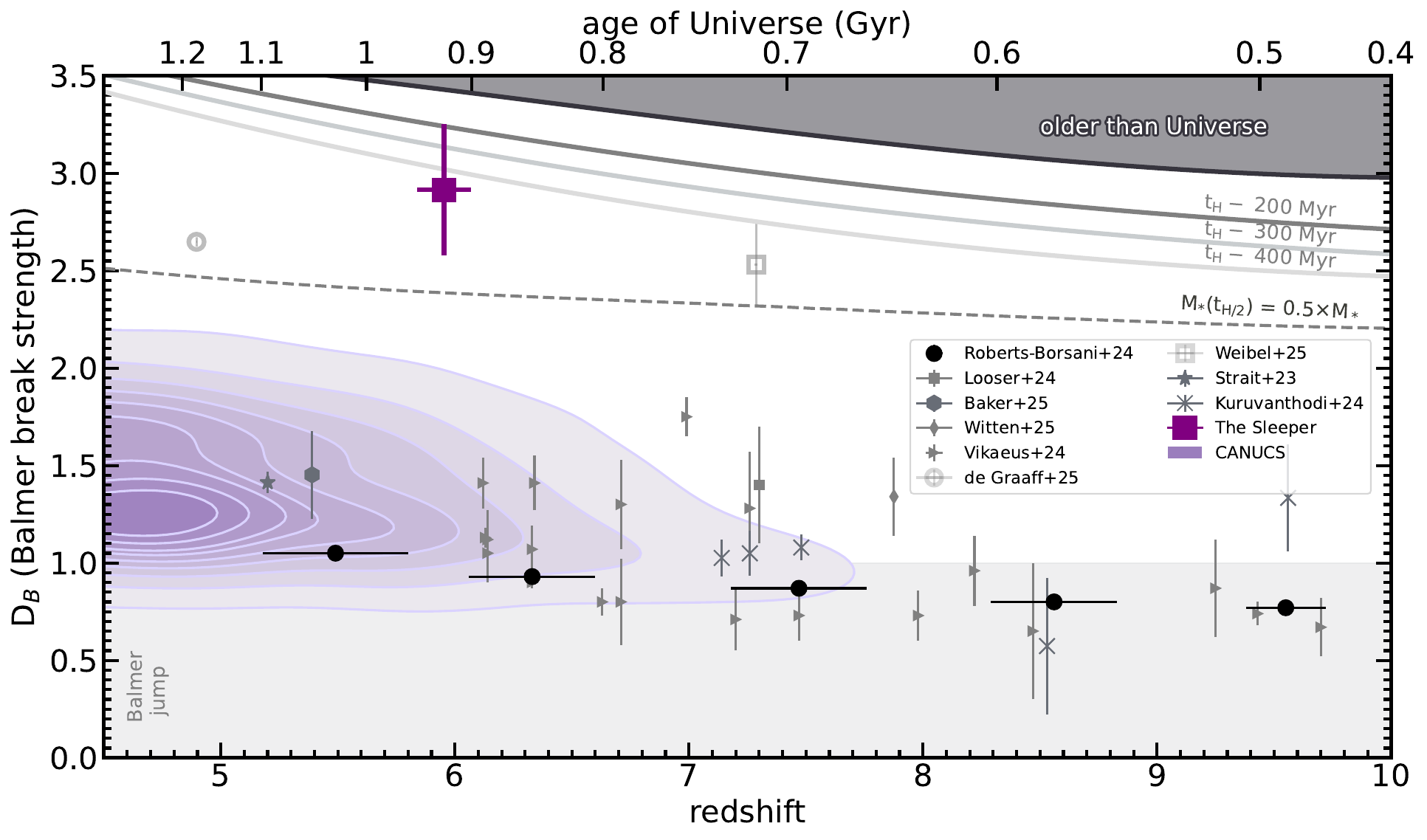}
\caption{Relationship between Balmer break strength ($\mathrm{D_B}$), age, and star formation history at $z \sim 5 - 10$. We show the break strength measured for a flux-limited CANUCS/Technicolor sample (shaded contours), spectroscopically-confirmed individual literature galaxies \cite{vikaeus2024, witten2025, strait2023, looser2024} (gray filled points), and stacks \cite{rb2024} (black filled points). We exclude ``little red dots," e.g., \cite{wang2024, setton2024}  because they may have a significant flux contribution from gas heated by AGN and not from stars, making the interpretation of their Balmer break strengths ambiguous. The solid lines show the evolution of maximally-old models: a stellar population with age equal to the \emph{Hubble} time, $t_H(z)$, and models assuming various onsets of star formation after the Big Bang ($200,~300,~\mathrm{and}~ 400$ Myr). These models assume an instantaneous burst (timescale, $\tau = 3$ Myr) and 10\% solar metallicity. The dashed line is for an extended burst model ($\tau = 400$ Myr at $z = 6$). Massive (log $M_\ast/M_\odot \geq 10$) quiescent galaxies at these redshifts \cite{degraaff2025_nature, weibel2025} are included as open points. The extreme Balmer break strength of \emph{The Sleeper} is best explained by a stellar population that formed at least 50\% of its stars in the first 500 Myr of cosmic history.} \label{fig:balmer_breaks} 
\end{figure} 

During the first billion years, where the young age of the Universe imposes a strict limit on the range of formation histories that can reproduce an observed SED, the detection of a strong Balmer break provides a particularly stringent test of the onset of star formation within $\Lambda$CDM \cite{steinhardt2024}. \edit{Importantly, the detection of such a pronounced break enables tight constraints on the past star formation of a galaxy by revealing the underlying evolved stellar populations}. We demonstrate this in Figure \ref{fig:balmer_breaks} by comparing the break strength of \emph{The Sleeper} with the expected $\mathrm{D_B}$ evolution of maximally-old stellar population models (Methods). In brief, we create a model assuming an instantaneous burst of star formation 200 Myr after the Big Bang followed by passive evolution, as well as an extended burst model where 50\% of the stellar mass is formed in half the \emph{Hubble} time at $z \approx 6$. The measured break strength lies between these two models, which implies that \emph{The Sleeper}'s photometry is best explained by a stellar population that formed at least half of its stars in the first 500 Myr. 

The vast majority of Balmer breaks discovered at $z \gtrsim 5$ belong to a population of $10^{7}-10^9~ \mathrm{M_\odot}$ galaxies observed during a temporary ($\lesssim  40$ Myr) \cite{dome2024} halt in-between bursts of star formation \cite{looser2024, witten2025, strait2023, vikaeus2024}, thereby showing weak to moderate breaks. In contrast, the strongest breaks observed at high redshift ($\mathrm{D_B} > 2$) --- typically associated with massive ($\mathrm{M_\ast \geq 10^{10}~ M_\odot}$) quiescent galaxies \cite{degraaff2025_nature, weibel2025} --- require $\gtrsim 200$ Myr of quiescence (Robbins et al. in prep). With a stellar mass of  log($M_\ast/M_\odot) = 9.14 \pm 0.09$ \edit{($12\times$ below the characteristic mass at $z \approx 6$ \cite{weaver2023})} and a remarkably strong Balmer break of $\mathrm{D_B} = 2.9_{-0.2}^{+0.3}$ \edit{that exceeds even those of canonical massive quiescent galaxies, the existence of \emph{The Sleeper} is unexpected. \emph{This discovery therefore occupies a new regime: a low-mass galaxy exhibiting evidence for an extended period of suppressed star formation at cosmic dawn.}}

\edit{The stellar population properties inferred from {\tt DENSE BASIS} support this interpretation.} The moderate equivalent widths of \emph{The Sleeper}'s rest-optical emission lines are consistent with a small remnant of surviving O-stars from the low levels of star formation: $W_0 (\OIII + \Hbeta)  = 251 \pm 20$~\AA~  and $W_0 (\Halpha + \NII) = 164 \pm 34$~\AA. These arise from the inferred moderate nebular conditions (ionization parameter, log$(U) \approx -3$) and chemical enrichment ($Z \approx 0.2 Z_\odot$) of the galaxy and are detected as flux excesses in the F335M and F460M filters, respectively. The instantaneous star formation rate (averaged over the past 10 Myr) is therefore constrained to SFR$_{10} = 0.58_{-0.37}^{+0.86}~\mathrm{M_\odot~yr^{-1}}$  by the nebular line flux measurements, indicating a rapid decline in recent star formation. Specifically, its rest-frame $\Halpha$ equivalent width is $\approx 0.8$ dex below the $\Halpha$ main sequence for $M_\ast > 10^8 M_\odot$ galaxies at $z \sim 6$ \cite{nc2024}. Furthermore, a decline in star formation is accompanied by a reddening and fading of the UV continuum and luminosity, in line with the measured properties of \emph{The Sleeper}  ($\beta = -1.58 \pm 0.46$ and M$_\mathrm{UV} = -17.32 \pm 0.44$). \edit{Together, these properties are consistent with an extended lull of star formation necessary to produce the extreme Balmer break of \emph{The Sleeper}}. 

 \edit{\emph{The Sleeper}'s star formation history can be accurately reconstructed owing to the extensive photometric sampling of key spectral features sensitive to different star formation timescales --- namely the UV continuum ($40 - 100$ Myr), the Balmer break ($0.2 - 1$ Gyr), and rest-optical emission lines ($4 - 10$ Myr)  \cite{iyer2025}. Its high redshift also reduces the allowable diversity of formation histories consistent with the observed SED, further tightening constraints on its past star formation.} Crucially, we use ``non-parametric" (flexible) models \cite{Leja2019}, as parametric forms are inadequate for modeling the episodic formation histories common at high redshift and often result in strongly biased stellar population parameters (e.g., \cite{carnall2019, zhang2025}). In fact, all parametric models yield poor fits to the data, with up to $\chi_{\nu}^2 = 3.3$ (see Methods). Following \cite{iyer2017}, we fit a model described by SFH $= (\mathrm{M_\ast, SFR_{100}, \{t_X\}})$, where ${t_X} = \mathrm{(t_{75}, t_{50}, t_{25}})$ are the lookback times at which the galaxy formed 75\%, 50\%, and 25\% of its stellar mass, respectively.

\begin{figure}[t]%
\centering
\includegraphics[trim={0.1cm 0.6cm 0.2cm 0.2cm},clip, width=\textwidth]{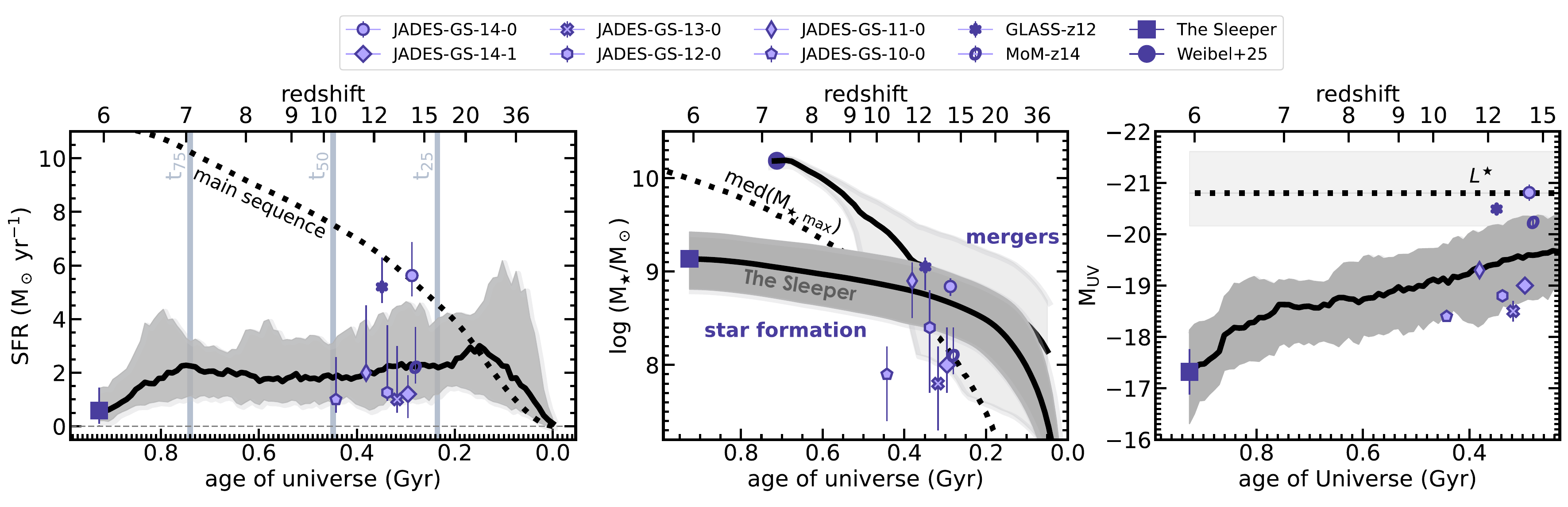}
\caption{\emph{The Sleeper} compared to spectroscopically-confirmed galaxies at $z > 10$ \cite{robertson2023, naidu2025, castellano2024}. Left: the median non-parametric star-formation history inferred using {\tt DENSE BASIS}. The shading indicates the 16th and 84th percentiles of the posterior estimated as a function of cosmic time. The vertical lines show when the galaxy assembled 25\%, 50\%, and 75\% of its stellar mass. \emph{The Sleeper} earns its name from the extended period of suppressed star formation at $z \lesssim 10$ relative to the \edit{time-dependent} main sequence \edit{traced by its stellar mass growth} \cite{speagle2014}. This results in a low current specific SFR (log $\text{sSFR}_{10} [ ~\textrm{yr}^{-1}] \approx -9.4$). Middle: the cumulative mass assembly history, accounting for mass loss due to stellar evolution. \emph{The Sleeper} forms a log($M_\ast/M_\odot) = 8.70_{-0.35}^{+0.32}$ progenitor at $z \sim 14$ in a rapid ($\approx 300$ Myr) burst, consistent with the stellar masses of UV-bright $z > 10$ galaxies. Right: the subsequent period of suppressed star formation results in a $\sim 2$ mag decline in the UV luminosity from $z \sim 14$ ($0.4\mathcal{L^{*}}$) to $z \sim 6$ ($0.04\mathcal{L^{*}}$), consistent with predictions from analytical models of  burstiness \cite{wyithe2014}. This assumes $\mathcal{L^{*}} = -20.75_{-0.81}^{+0.64}$ \cite{willott2024}. }\label{fig:sfh}
\end{figure} 

The median star formation and cumulative mass assembly histories are shown in Figure \ref{fig:sfh} (see Methods for the maximum-likelihood star-formation history). They reveal a rapid rise in star formation at $z \gtrsim 17$, during which the galaxy formed 25\% of its total stellar mass, followed by steady growth at $\sim 1-4 ~\mathrm{M_\odot~ yr^{-1}}$ for $\Delta t \approx 500$ Myr and a factor of SFR$_{100}$/SFR$_{10} \approx 2$ decline over the past 100 Myr. This implies a short duty cycle, i.e., the fraction of time spent on or above the star forming main sequence \cite{speagle2014} relative to the lifetime of the galaxy, of $f_\mathrm{duty} = 16\%$, compared with $f_\mathrm{duty} \gtrsim 40\%$ for similar-mass galaxies at $z > 5$ \cite{dome2024, cole2025}. The star formation history is consistent with the moderate star formation rates of several spectroscopically-confirmed galaxies at $z \sim 10 - 14$ \cite{robertson2023, castellano2024, naidu2025, carniani2024}. Likewise, \emph{The Sleeper} forms a log($M_\ast/M_\odot) = 8.70_{-0.35}^{+0.32}$ progenitor at $z \sim 14$, in agreement with the inferred stellar masses of the aforementioned. \edit{\emph{The Sleeper}'s assembly history therefore brackets the range of plausible stellar masses of relics from the first 500 Myr, providing an empirical benchmark for identifying similar ancient systems in deep-field surveys.}

\edit{The rapid, early assembly and short duty cycle of \emph{The Sleeper} indicate either bursty star formation within the first $\approx 300$~Myr or enhanced star formation efficiencies at high redshift relative to local values ($<5\%$ for a dark matter halo mass of $\log(M_h/M_\odot) \simeq 10 - 11$ \cite{baldry2008}). Both mechanisms have been proposed to explain the  overabundance of UV-luminous galaxies at $z>10$  \cite{sun2023, franco2025}. As a relic, \emph{The Sleeper} directly tests these scenarios. 
Figure \ref{fig:sfh} shows that \emph{The Sleeper}'s short-lived, active phase occurred early ($z \gtrsim 17$), consistent with a feedback-regulated starburst that rapidly consumed or expelled a large fraction of its gas reservoir \cite{ferrara2024_z14}. Although the inferred star formation history shows a single, dominant burst at this epoch, the $\approx 100$ Myr temporal resolution from SED fitting \cite{iyer2019} precludes resolving the $< 5$ Myr fluctuations seen in simulations. The early burst is therefore likely an average of multiple short events. The resulting gas depletion suppresses \emph{The Sleeper}'s star formation for several 100 Myr, producing the pronounced Balmer break observed at $z \approx 6$.} Our comparison with the \textsc{Sphinx}$^{20}$, \textsc{thesan}, and \textsc{Flares} simulations supports this picture. We find that the strongest Balmer breaks at $\sim 10^9 ~\mathrm{M_\odot}$ arise in simulations with a large scatter in burst intensity, $\sigma_\eta > 0.4 $ dex, where $\eta = \mathrm{log_{10}(SFR_{10}/SFR_{100})}$ (Methods). This agrees with semi-analytical models requiring moderate to large stochastic scatter to match the observed UV luminosity function at $z > 10$ (e.g., \cite{franco2025}).  

\edit{\emph{The Sleeper}’s early mass assembly also constrains the allowed star formation efficiency at cosmic dawn. Its mass growth at $z \gtrsim 14$ exceeds the median stellar mass, $\log(M_\star/M_\odot) = 8.17 \pm 0.45$, of the most massive galaxy expected in the CANUCS/Technicolor survey} ($\approx 29$ sq. arcmin), yet remains consistent with $\Lambda$CDM within $2\sigma$ in an extreme value statistics framework (Methods). We adopt a baryon-to-stellar conversion efficiency peaking at $\epsilon_\star \simeq 14\%$, consistent with the stellar-halo mass relation \cite{sun2016} and obviating the need for abnormally high efficiencies ($>30\%$) proposed to explain UV-bright galaxies at this epoch \cite{harikane2023}. \edit{Only at the extreme tail ($3\sigma$) of the predicted stellar mass distribution ---$\log M_\star \simeq 9.3$, hosted by haloes of $\log(M_h/M_\odot) \simeq 11$ \cite{sun2016}---would the implied efficiency reach $\sim40$--$60\%$ (see Methods). \emph{The Sleeper} therefore provides empirical evidence for models in which UV-bright phases arise from burstiness \cite{sun2023} rather than a redshift-dependent increase in efficiency \cite{donnan2025_efficiency}.}

\begin{figure}[ht!]%
\centering
\includegraphics[trim={0.21cm 0.2cm 0cm 0.4cm}, clip, scale=0.328]{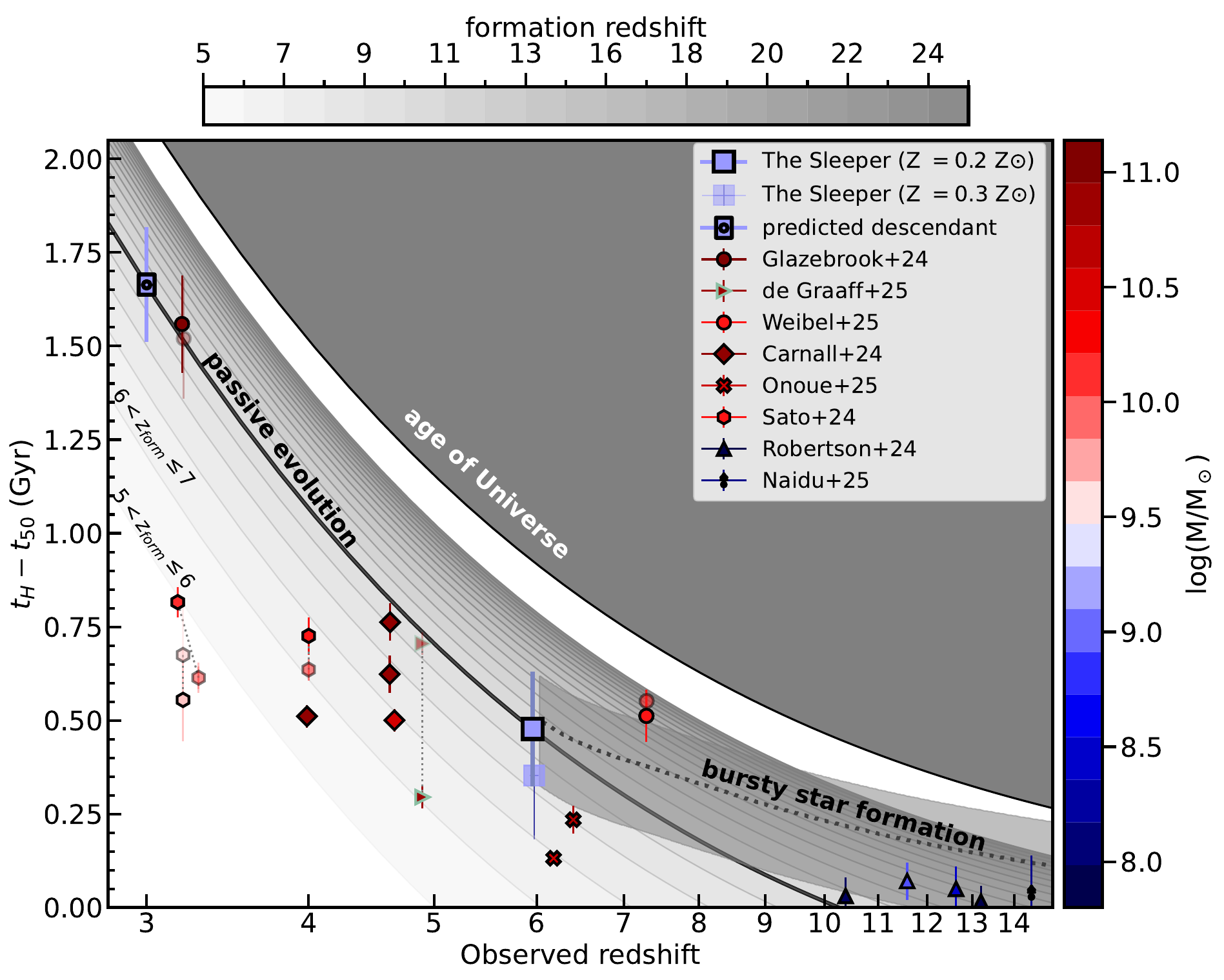}
\caption{Evolutionary pathways of the oldest galaxies observed in the first two billion years. We show the lookback time at which the galaxy formed 50\% of its stellar mass as a function of observed redshift for spectroscopically-confirmed massive quiescent galaxies at $z \gtrsim 3$ \cite{glazebrook2024, degraaff2025_nature, weibel2025, carnall2024, onoue2025, sato2024} and UV-luminous galaxies at $z > 10$ \cite{robertson2023, naidu2025}. Lighter symbols use the $t_{50}$ corresponding to a higher metallicity solution, where available. Shaded tracks show the ages of passively evolving simple stellar populations at $5 < z_\mathrm{form} \leq 25$. The star formation history and stellar mass of \emph{The Sleeper} are broadly consistent with the measured ages of the first galaxies at $z \sim 14$. Given that the oldest ($t_{50} \gtrsim 500$ Myr) galaxies observed at these redshifts tend to be extremely massive ($\gtrsim 10^{10} ~ \mathrm{M_\odot}$), \emph{The Sleeper} presents a unique opportunity to study a possible progenitor of the low-mass quiescent population at $z \lesssim 3$. Assuming passive evolution at its current star formation rate, \emph{The Sleeper} predicts a population of ancient, low-mass galaxies at $z \sim 3$ which are missing from current samples likely due to observational biases \cite{cutler2024}. }\label{fig:age_evolution}
\end{figure} 
\edit{The predicted UV luminosity decline from theory \cite{wyithe2014, gelli2025} following the early, burst-driven growth phase supports the interpretation that \emph{The Sleeper} had a transient UV-bright phase at $z \gtrsim 10$}. We simulate its UV luminosity evolution by evolving a mock stellar population from $z \approx 15$ to  $z \approx 6$ based on its star formation and mass assembly history. The resulting prediction matches both the reported UV luminosities of several $z > 10$ sources and the measured value of $\mathrm{M_{UV}} \approx -17.3$ ($0.04\mathcal{L^{*}}$) for \emph{The Sleeper}. Figure \ref{fig:sfh} therefore summarizes our key result: some fraction of the luminous population at $z > 10$ can be explained by a rapid ($\approx 300$ Myr) and efficient ($\epsilon_\star \approx 14\%$) period of bursty star formation, followed by a period of much slower growth marked by suppressed star formation and a corresponding ($5 \times $) decline in the UV luminosity, \edit{resulting in a Sleeper-like relic}. 

Finally, we consider possible evolutionary paths of \emph{The Sleeper} --- and by extension, its progenitors --- following its observation at $z = 5.95 \pm 0.06$. As we show in Figure \ref{fig:age_evolution}, the inferred mass--weighted age of \emph{The Sleeper} ($480 \pm 10$ Myr) in a Universe that is only a billion years old places it as one of the most ancient galaxies yet identified. Its mass assembly history demonstrates that UV-bright galaxies at cosmic dawn need not evolve into massive quiescent galaxies at $z > 3$. If the current low star formation rate of \emph{The Sleeper} ($\mathrm{SFR_{10} \approx 0.6~M_\odot~yr^{-1}}$) continues to decline, it will evolve passively into an ancient ($\approx 1.6-1.9$ Gyr old) dwarf ($\approx 0.02 \mathcal{M}^*$, assuming a characteristic stellar mass of log$(\mathcal{M}^*/M_\odot) = 10.83$ \cite{weaver2023}) at $z = 3$. Figure \ref{fig:age_evolution} illustrates that the predicted descendant population is missing from current samples \cite{cutler2024}, even in higher density environments \cite{pan2025}, which tend to host evolved stellar populations. \edit{This is surprising given that most deep-field surveys are mass-complete for the predicted descendant, which at $m_\mathrm{F150W} = 28.2$ AB would be sufficiently bright to be detected in current \emph{JWST} extragalactic fields. Therefore, the dearth of this descendant population may be due to selection biases. Alternatively, the overabundance of Sleeper-like objects relative to simulations (see Methods) could imply that other mechanisms, such as galaxy-galaxy interactions or rejuvenation (e.g., \cite{witten2025}) may be at play in shaping their future growth.}  

The discovery of \emph{The Sleeper} thus reveals an alternative evolutionary pathway to the canonical merger-driven growth of early luminous galaxies. It presents striking observational evidence that some fraction of these evolve into low-mass descendants rather than the most massive systems.  Future spectroscopy will confirm the properties of this unique system and test whether galaxies like \emph{The Sleeper} represent a transient phase or a dominant, yet hidden, population of distant fossils. 

\section{Methods}\label{sec11}
\subsection{Cosmology and Conventions}
For this work, we assume a flat $\mathrm{\Lambda CDM}$ cosmology with $\Omega_{\rm M} = 0.3$, $\Omega_{\rm \Lambda} = 0.7$, and $\mathrm{H_0 = 70\; km\; s^{-1}\; Mpc^{-1}}$ when calculating physical parameters and their evolution over cosmic time. We quote magnitudes in the AB system \cite{oke_gunn_1983}, where AB $ = - 2.5~\mathrm{log_{10}}(f_\nu/\mathrm{Jy}) + 8.90$. All reported values and uncertainties are medians and 68\% \edit{credible intervals}. We use vacuum wavelengths for emission lines, and report equivalent widths in the rest-frame. \edit{The stellar libraries we use assume Chabrier  \cite{chabrier2003}, Kroupa \& Boily\cite{kroupa2002}, or Salpeter \cite{salpeter1955} initial mass functions. Conversions to Chabrier are $-0.05$ dex for Kroupa \& Boily and $-0.24$ dex for Salpeter in both stellar mass and star formation rate \cite{speagle2014}.} 
\label{sec:conventions}

\def\arraystretch{1.3}
\begin{table}
    \centering
    \caption{Summary of stellar population properties of \emph{The Sleeper}.}
    \begin{tabular}{l c}
    \hline
    \hline
ID & CANUCS-A370-2228423   \\
RA[ICRS]  &  02:40:11.39    \\
DEC[ICRS]   & -01:36:35.74   \\
\hline
redshift & $5.95\pm0.06$  \\
$\log_{10}(M_\mathrm{star} / \textrm{M}_\odot)$$^\star$  & $9.14\pm0.09$\\ 
$\text{D}_\text{B}$ &  $2.9_{-0.21}^{+0.34}$   \\
mass-weighted age [Myr] & $480 \pm 10$  \\
Circularized effective radius ($r_\mathrm{eff}$) [pc] & $643\pm52$ \\
UV slope $\beta$ & $-1.58 \pm 0.46$  \\
$M_\mathrm{UV}$  & $-17.32\pm 0.44$ \\ 
$\text{SFR}_{100}~[\textrm{M}_\odot \,\textrm{yr}^{-1}$]& $1.03_{-0.50}^{+1.31}$ \\ 
$\text{SFR}_{10}~[\textrm{M}_\odot \,\textrm{yr}^{-1}$]& $0.58^{+0.86}_{-0.37}$ \\ 
log $\text{sSFR}_{10}~[\textrm{yr}^{-1}$]& $-9.37_{-0.36}^{+0.31}$ \\
$A_\mathrm{V}$ [mag]&  $0.18_{-0.13}^{+0.23}$   \\
$W_0$ (\OIII + \Hbeta) [\AA] &  $251 \pm 20$   \\
$W_0$ (\Halpha + \NII) [\AA] &  $164 \pm 34$   \\

    \hline
    \end{tabular}
    \label{tab:summary}
\end{table}

\subsection{HST and JWST Observations}\label{sec:observations}
Our analysis is based on multi-wavelength imaging from the \emph{JWST in Technicolor} Cycle 2 GO Program \#3362 (PI: A. Muzzin), a followup survey targeting three well-known strong lensing clusters observed in the CANUCS NIRISS Cycle 1 GTO Program \#1208 \cite{willott2022}: Abell 370, MACS J0416.1-2403, and MACS J1149.5+2223. These surveys operated NIRCam and NIRISS in parallel, hence each cluster field is accompanied by a NIRCam flanking field. Our target lies in the flanking field of Abell 370, which was observed in all of the broad and medium band filters, as well as two narrow bands (F164N and F187N). The exposure times are $5.7 - 10.3$ ks for each field, reaching SNR $= 3$ in a $0.3$\arcsec~ diameter aperture for a $m_{\mathrm{AB}}= 28.8 - 30.5$ point source \edit{in the medium and broad bands}. We supplemented the NIRCam imaging with archival \emph{HST} imaging from the Hubble Frontier Fields program \cite{lotz2017} and HST-GO-16667 (PI: M. Bradac). The combined dataset consists of deep imaging in $27 - 29$ filters over $29.1$ arcmin$^2$ across the three flanking fields, providing exquisite photometric sampling of distant galaxies at $0.9 - 4.4~\mu$m. 

The CANUCS and Technicolor survey designs as well as details on data reduction, point spread function homogenization, source detection, and the photometry pipeline are described in \cite{noirot2023, willott2024, asada2024, sarrouh2024}. In brief, the raw data was reduced using the public grism redshift \& line analysis software {\tt Grizli} \cite{brammer2019} and standard data reduction procedures \cite{kokorev2022}, including masking artifacts and astrometric calibration to the Gaia Data Release 3 catalog \cite{gaiadr3}. The CANUCS pipeline implements custom procedures for bad pixel masks, persistence flagging, and 1/$f$ noise removal, detailed in the survey paper \cite{sarrouh2025}. 

\begin{figure}[t]%
\centering
\includegraphics[trim={0.2cm 0.1cm 0.1cm 0.1cm}, clip, scale=0.148]{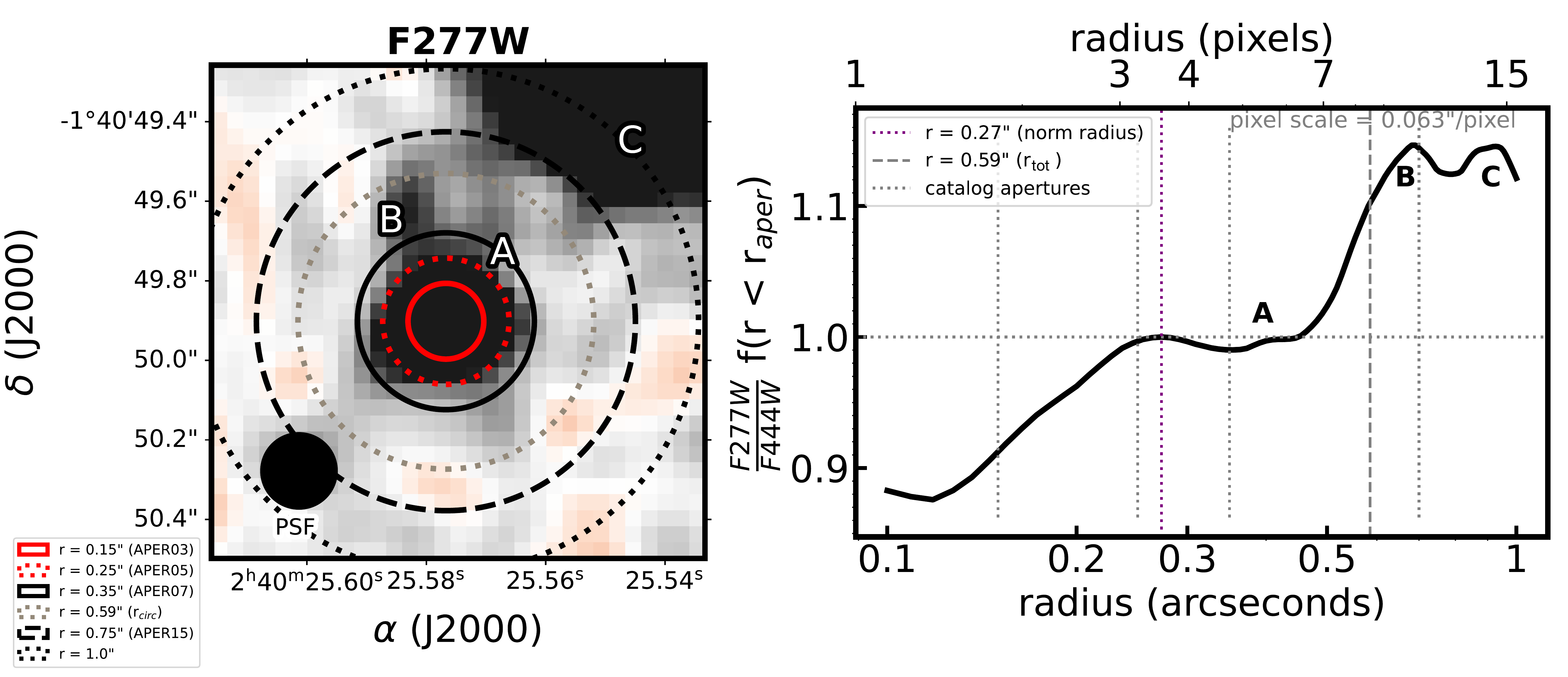}
\caption{The flux profile of \emph{The Sleeper} . The left panel shows a $\sim 1\times1$\arcsec~ cutout from the PSF-homogenized F277W image in logarithmic scaling, showing the 2D flux profile of its companions labelled B (CANUCS-A370-2217627, $z = 6.78 \pm 0.14$) and C (CANUCS-A370-2201097, $z = 6.00 \pm 0.03$). The catalog apertures are also shown. Right panel: 1D flux profile of \emph{The Sleeper}  determined using the ratio of its growth curves in the highest S/N filter (F277W) and the PSF reference (F444W), normalized at $r = 0.27$\arcsec. While the scaled Kron aperture, $r_\mathrm{tot} = 2.5\times r_\mathrm{circ}$, contains the total source flux, it also includes $\sim 10\%$ contamination from B and C.  This would introduce a systematic of the same magnitude in estimating its photometric redshift and stellar population parameters. Therefore, we determine its total flux using the optimal aperture, $r = 0.25$\arcsec,  scaled to $r=0.35$\arcsec~.}\label{fig:photometry}
\end{figure} 

\subsection{Photometry of \emph{The Sleeper} }\label{sec:photometry}
We obtained photometric measurements using {\tt photutils} on PSF-homogenized images in circular apertures of $r=0.15$\arcsec, 0.25\arcsec, 0.35\arcsec, 0.75\arcsec, 1.0\arcsec, and $r_\mathrm{tot} = 2.5\times r_\mathrm{circ}$, the circularized Kron radius \cite{kron1980}. This scaled Kron aperture typically encloses $> 96\%$ of the source flux. However, the exact fraction is based on the source's light profile, which can vary significantly due to morphology and the presence of a companion. \emph{The Sleeper} is separated by $\sim 0.5$\arcsec~ from a neighbor at $z = 6.00 \pm 0.03$ that is $\approx 2$ magnitudes brighter than \emph{The Sleeper} ($m_\mathrm{AB} = 27.7$ in F277W). In Figure \ref{fig:photometry}, we show the flux profile of \emph{The Sleeper} computed using the ratio of its growth curves in F277W, the highest S/N filter, and F444W, the PSF reference. $\sim 10\%$ of the source flux in the total aperture, $r_\mathrm{tot} = 0.59$\arcsec~, is due to contamination from the neighboring object. This exceeds the typical flux systematic of $3-5\%$ (e.g., \cite{brammer2008}) that is added to photometry for stellar population modeling and which faint sources such as \emph{The Sleeper}  are particularly sensitive to.

For this reason, we choose $r = 0.25$\arcsec~ as the optimal aperture. It is small enough that it maximizes the S/N in the largest number of filters and large enough that it captures most of its light without contamination from the companion. The latter is also important for minimizing systematics from residual astrometric or PSF matching uncertainties. Following \cite{whitaker2011}, we measure the total flux of \emph{The Sleeper} by scaling the flux in the optimal aperture to that in the $r = 0.35$\arcsec~ aperture, with the uncertainties scaled accordingly. This aperture correction is on the order of $\sim 1\%$, with a factor of $\sim 2$ increase in the flux uncertainties. A local sky subtraction is applied to these fluxes as well as a correction for Galactic dust extinction along the line of sight assuming \cite{fitzpatrick1999} and a total-to-selective extinction ratio $R_\mathrm{V} = A_V/E(B-V) = 3.1$, with $E(B-V)$ values from \cite{schlafly_finkbeiner2011}.  We perform all further analysis on this optimal photometry. 

\subsection{Empirical Properties}\label{sec:empirical} 
Using {\tt Galfit} \cite{peng2010}, we measure the effective radius, $r_e = 0.11 \pm 0.01$\arcsec~ in the JWST/NIRCam F277W filter assuming a single S{\'e}rsic component with index, $n = 1.4 \pm 0.3$, \edit{consistent with an exponential disk}. This corresponds to $r_\mathrm{eff} = 643 \pm 52$ pc at $z = 5.94$. For comparison with UV-bright galaxies at cosmic dawn, we also measure the rest-UV slope, $\beta_\mathrm{UV} = -1.58 \pm 0.46$, by assuming an intrinsic stellar continuum described by a power law $f_\lambda \propto \lambda^{\beta_\mathrm{UV}}$ at $\lambda_\mathrm{rest} < 3000$ \AA, corresponding to $\lambda_{obs} < 2.1~\mu$m. Due to the high photometric sampling of the CANUCS/Technicolor data, 12 filters are used in the fit. Additionally, we estimate the absolute UV magnitude, $M_\mathrm{UV} = -17.32 \pm 0.44$, using the rest-frame filter closest to $1500$ \AA~ in which the galaxy is detected at S/N $ \gtrsim 2$. At $z = 5.95$, this corresponds to \emph{HST}/ACS F814W, where we measure $f_\nu = 6.55 \pm 2.68~$ nJy. As a result of the robustly-constrained photometric redshift, owing to the presence of both the Lyman and Balmer breaks, the uncertainties on $M_\mathrm{UV}$ are dominated by the uncertainties on the F814W flux. We report these empirical properties in Table \ref{tab:summary}. 

\subsection{Spectral Energy Distribution Fitting}\label{sec:sed_fitting}
We estimated the physical properties of \emph{The Sleeper}  using three spectral energy distribution (SED) fitting codes. This approach allows us to explore the impact of differences in stellar population libraries, star formation history (SFH) parameterizations, and minimization algorithm on the derived quantities. \edit{As these stellar libraries incorporate different isochrones and IMF assumptions, these are thereby implicitly tested.} In all cases, we assume that the observed SED is dominated by emission from stars and ionized gas. This is reasonable given that we do not find any evidence of point-like emission from an underlying active galactic nucleus (AGN) based on continuum-subtracted images. Additionally, \emph{The Sleeper}  is spatially-resolved and shows an extended morphology (Sect. \ref{sec:empirical}) \edit{and therefore would not be selected as an AGN or ``little red dot" based on compactness criteria \cite{akins2025}}. 

\subsubsection{eazy-py}\label{sec:eazy}
For reproducibility, the public CANUCS catalogs include physical properties derived with {\tt eazy-py}, a Python version of the widely-used EAZY photo-z code \cite{brammer2008}. It derives these quantities by fitting a non-negative linear combination of user-defined stellar population templates, with the redshift and scaling of each template as free parameters. We use a set of 15 spectra based on the templates from \cite{larson2023}, which augment the standard set ({\tt tweak\_fsps\_QSF\_v3}) with templates that are more representative of the highly star-forming, younger, and less dusty galaxies at high redshift \cite{boyett2024, papovich2025}. Three of these are custom templates where the \OIII$\lambda \lambda 4959, 5007$ doublet is boosted by a factor of 3 to capture the most extreme emitters at $z > 4$. The redshift is allowed to vary from $z = 0-20$ and we apply a luminosity prior which assigns low probabilities to high redshift solutions for bright galaxies. Following this procedure, we derive $z = 5.94 \pm 0.04$ with {\tt eazy-py}, which we use as a redshift prior with {\tt DENSE BASIS}. 

\subsubsection{Dense Basis}\label{sec:dense_basis}
\begin{figure}[t!]%
\centering
\includegraphics[trim={0.2cm 0.1cm 0.1cm 0.1cm}, clip, scale=0.372]{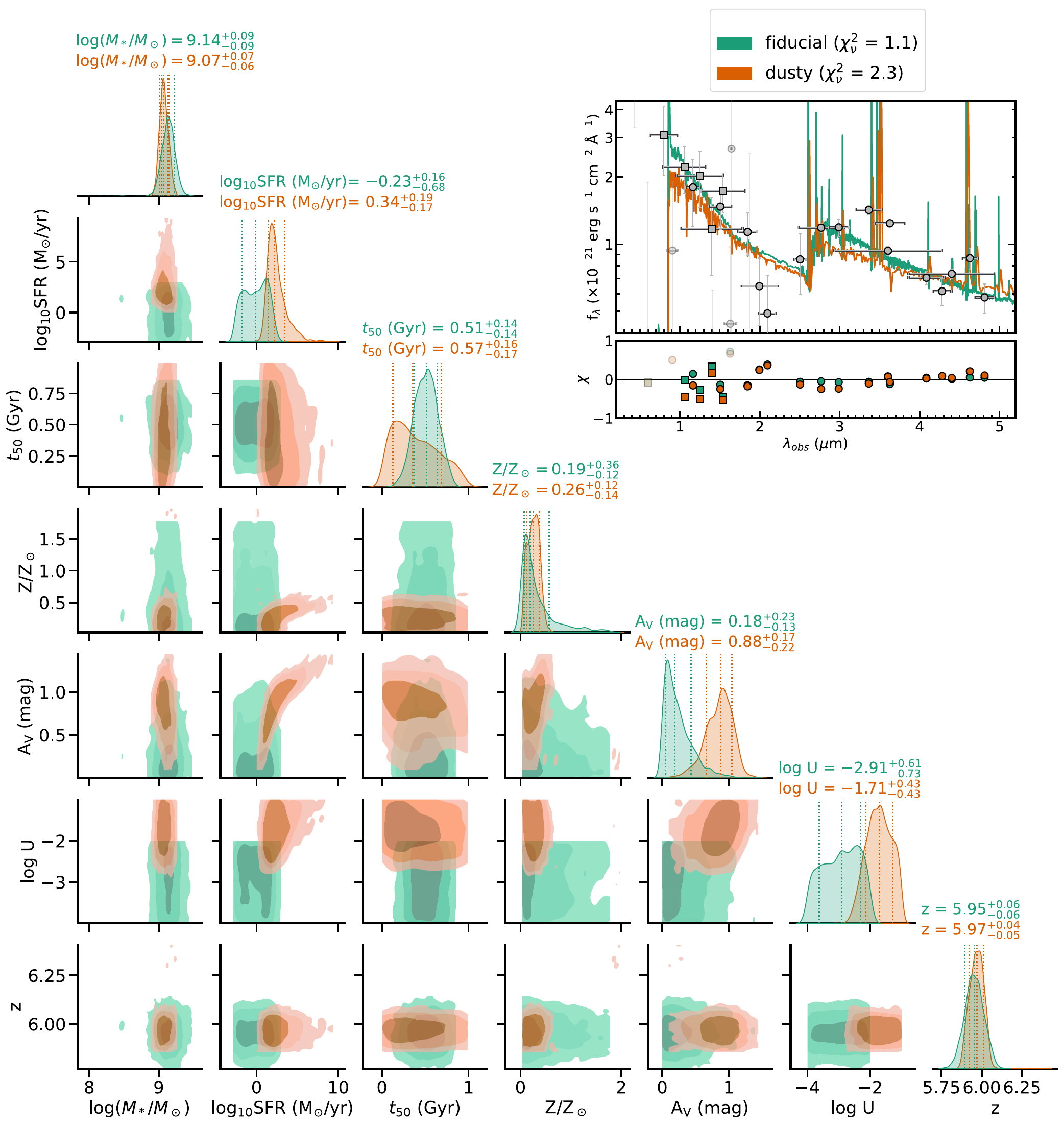}
\caption{A comparison of the covariant posteriors and SEDs for our fiducial fit from {\tt DENSE BASIS} ($\chi^{2}_\nu = 1.1$) and a dusty solution from Bagpipes ($\chi^{2}_\nu = 2.3$). The residuals, $\chi$, between the model photometry and measured fluxes suggest that that flux excess in F300M relative to F250M arises from the stellar continuum of a population dominated by evolved B stars, rather than nebular emission lines. The residuals are restricted to $\chi = \pm 1$ to highlight the excellent fit of the fiducial model, however we note that the dusty model has up to $\sim 7\sigma$ offsets in several rest-UV filters, \edit{with the largest occuring in F435W, F070W, F814W, and F090W}. Contours bound the 68\%, 95\%, and 99.7\% regions of the likelihood. In both cases, we use a non-parametric SFH reconstructed using Gaussian processes \cite{iyer2017, iyer2019}. The stellar mass posteriors account for mass loss due to stellar evolution. }\label{fig:corner}
\end{figure}

{\tt DENSE BASIS} \cite{iyer2017} permits a more flexible stellar population model than is possible with {\tt eazy-py}. This framework allows for ``nonparametric" (multi-parameter) star formation histories, which are better suited for capturing multiple bursts of star formation --- found to be  ubiquitous in high redshift galaxies (e.g.,\cite{cole2025}) --- and maximally old stellar populations \cite{Leja2019}. The SFH is reconstructed using Gaussian processes \cite{rw2005} by splitting the total mass formed into $N$ time bins of equal mass with the normalization determined by the instantaneous SFR. This formulation mitigates the limitations of the continuity model in {\tt Prospector} \cite{johnson2021} by permitting more dramatic changes in the SFH such as sudden bursts and quenching. We use $N=3$, parameterizing the age of the Universe at which the galaxy formed 25\%, 50\%, and 75\% of its stellar mass ($t_{25}, t_{50}, ~\mathrm{and}~ t_{75}$, respectively), as this has been found to adequately recover episodic star formation in semi-analytic models \cite{iyer2019}.  {\tt DENSE BASIS} uses \textsc{Fsps} \cite{conroy_gunn2010} with the MILES spectral library \cite{vazdekis2010}, nebular emission lines computed with the CLOUDY photoionization code \cite{ferland1998}, and intergalactic (IGM) absorption using the model of \cite{madau1995}. We assume the Calzetti starburst law, as the attenuation curves of high redshift galaxies are generally flat \cite{markov2025_nature}. The gas-phase and stellar metallicities are coupled, dust is allowed to vary freely in the range  $A_V \in$ [0, 4] mag with an exponential prior, and the remaining parameters vary uniformly in the following ranges: $Z \in$ [0.03$Z_\odot$, 1.8$Z_\odot$], $M_\ast \in$ [$10^{7} M_\odot$, $10^{12}M_\odot$], log(sSFR) $\in$ [$-7$, $-14$], log $(U)\in$ [$-4$, $-2$]. The observed SED of \emph{The Sleeper} is best reproduced by this model ($\chi^2_\nu = 1.14$), therefore, we use this as the fiducial. We show the estimated posteriors in Figure \ref{fig:corner}, \edit{which are smooth, unimodal, and free from boundary effects,} demonstrating that the \edit{stellar population properties are well constrained under the adopted priors}. The posterior medians and credible intervals from {\tt DENSE BASIS} are listed in Table \ref{tab:summary}. 

\subsubsection{Bagpipes}\label{sec:bagpipes}
Finally, we fit the photometry using the Bayesian Analysis of Galaxies for Physical Inference and Parameter EStimation (\textsc{Bagpipes}) software \cite{carnall2018}. We use the same non-parametric SFH parameterization and a similar setup of priors as with {\tt DENSE BASIS}. \textsc{Bagpipes} offers two stellar library options: the 2016 version of BCO3 \cite{bc03} and BPASS v2.2.1 \cite{bpass2009}, which assume a Kroupa \& Boily and Salpeter IMF, respectively. Although BC03 produced a reasonable fit ($\chi^2_\nu = 1.7$), the ionization parameter converged towards the edge of the prior space (log $(U) = -2$), suggesting the need for bluer models. Therefore, we use BPASS, which accounts for binary star evolution, thereby predicting a higher ionizing flux than BC03 and allowing us to explore stellar populations with ionization parameter log $(U)\in$ [$-2$, $-1$]. We perform sampling using the \edit{\textsc{PyMultiNest} nested sampling algorithm \cite{buchner2014}}. 

There is remarkable agreement between the \textsc{Bagpipes} and {\tt DENSE BASIS} redshift and stellar mass solutions ($\Delta z = 0.02$ and $\Delta \mathrm{log}(M_\ast/M_\odot) = 0.07$ dex, respectively), \edit{especially given that we do not implement a redshift prior on the former}. \textsc{Bagpipes} reproduces the observed SED with $\approx 1$ mag of dust extinction, compared to the modest $\approx 0.2$ mag in the fiducial model from {\tt DENSE BASIS}. This is expected due to the differences in the underlying stellar libraries, as the bluer BPASS models require more dust to match the red UV slope ($\beta \approx -1.6$). Although 1 mag of extinction can boost the Balmer break strength by up to D$_\mathrm{B} \sim 0.2$ for stellar populations younger than 1 Gyr \cite{d'eugenio2021}, the \textsc{Bagpipes} model exhibits a weaker Balmer break ($\mathrm{D_B = 1.8}$) than the fiducial (Figure \ref{fig:corner}). This arises from a higher nebular continuum around rest-frame $3500$~\AA~and mass transfer between binaries, resulting in more hot stripped stars and fewer red supergiants in BPASS \cite{ma2018}. While the \textsc{Bagpipes} model produces a poorer fit to the data ($\chi^2_\nu = 2.3$), it is not ruled out due to the large uncertainties on F250M (S/N $= 3.3$). It is also important to note that the observed SED depends on the dust geometry, and a strong Balmer break can result from differential dust between the stellar and nebular components \cite{faisst2024}. Spectroscopy or spatially-resolved observations are needed to investigate this dusty scenario. 

\subsubsection{Impact of Dust}\label{sec:dust}
The dust prior and law can dramatically impact the inferred stellar population, therefore we investigate the impact of our choices on our SED fitting results. We employ an exponential dust prior (with width, $\sigma_\mathrm{A_V} = 0.3$) in our fiducial fit (Sect. \ref{sec:dense_basis}) because \edit{there is evidence that a} flat prior may overestimate dust for faint galaxies with intrinsically low-to-moderate extinction \cite{pacifici2023}. With {\tt DENSE BASIS}, a flat dust prior yields A$_\mathrm{V} = 0.34^{+0.44}_{-0.26}$, with a poorer fit to the data ($\chi^2_\nu = 1.8$). Additionally, we test using a steeper dust law than Calzetti. While high redshift galaxies generally have flat dust laws, at $z > 2$, this is not the case for galaxies with old stellar populations, low sSFRs, and low dust optical depths \cite{markov2025}. Indeed, \textsc{Bagpipes} fits with an SMC dust law yield improved fits to the data, with A$_\mathrm{V} = 0.36^{+0.11}_{-0.11}$ ($\chi^2_\nu = 1.7$) and A$_\mathrm{V} = 0.36^{+0.11}_{-0.09}$ ($\chi^2_\nu = 1.9$) for BC03 and BPASS, respectively. In summary, the data strongly favors solutions with low-to-moderate dust (A$_\mathrm{V} \approx 0.2 - 0.4$).  

\subsection{Star Formation History Testing}\label{sec:sfh} 
Here, we compare our fiducial fit from {\tt DENSE BASIS} with models generated using parametric star formation histories. We test six widely-used functional forms implemented within \textsc{Bagpipes}: double power law, exponentially declining (``tau" models), delayed exponentially declining, lognormal, burst, and constant star formation histories. A major limitation of these is that they are strongly biased against early star formation, hence tend to favor younger stellar populations \cite{carnall2019}. Additionally, unlike with non-parametric models, it is not possible to explicitly set physically-motivated priors on fitted parameters.  The parametric SFHs all produce significantly underfitted models, with up to $\chi^2_\nu = 3.3$. In particular, they produce a UV slope that is too red and/or overpredict the strength of the Balmer break. Although the $\tau$, delayed$-\tau$, and lognormal SFHs produce reasonable fits ($\chi^2_\nu \sim 2$), the posteriors of their fitted parameters are multimodal, \edit{consistent with degenerate solutions}. This implies that, \edit{excluding the lognormal and $\tau$ models, simple parametric forms} are not flexible enough to capture recent quenching \cite{suess2022} or multiple bursts of star formation \cite{haskell2024} in \emph{The Sleeper}.  %

\subsection{Measurement of D$_\mathrm{B}$ and its Evolution}\label{sec:db_evolution}
Using the definition of \cite{rb2024}, we measure the Balmer break strength, $D_B = F_{\nu, 4225}/F_{\nu, 3565}$ from the best-fit SED using two emission line-free windows covering the red and blue portions of the stellar continuum, respectively: $\lambda_\mathrm{rest} = 3500 - 3630~ \mathrm{Å}$ and $4160 - 4290~ \mathrm{Å}$. By this definition, a galaxy with a Balmer \emph{break} has $D_B > 1$ and that with a Balmer \emph{jump} or \emph{inverse} break has $D_B < 1$. The uncertainties on $D_B$ are determined by interpolating between those of the observed photometric bands flanking the red and blue windows. We compare the Balmer break strength of \emph{The Sleeper}  with those of spectroscopically-confirmed sources in the literature  \cite{weibel2025,  vikaeus2024, looser2024, strait2023, degraaff2025_nature, kuruvanthodi2024, baker2025}. We use their reported $D_B$ measured using the same windows as above. Where not available (i.e., for \cite{strait2023, degraaff2025_nature, kuruvanthodi2024, baker2025}), we estimate $D_B$ from their spectra using our chosen definition.

\begin{figure}[t]%
\centering
\includegraphics[trim={0.1cm 0.1cm 0.1cm 0.2cm}, clip, scale=0.4]{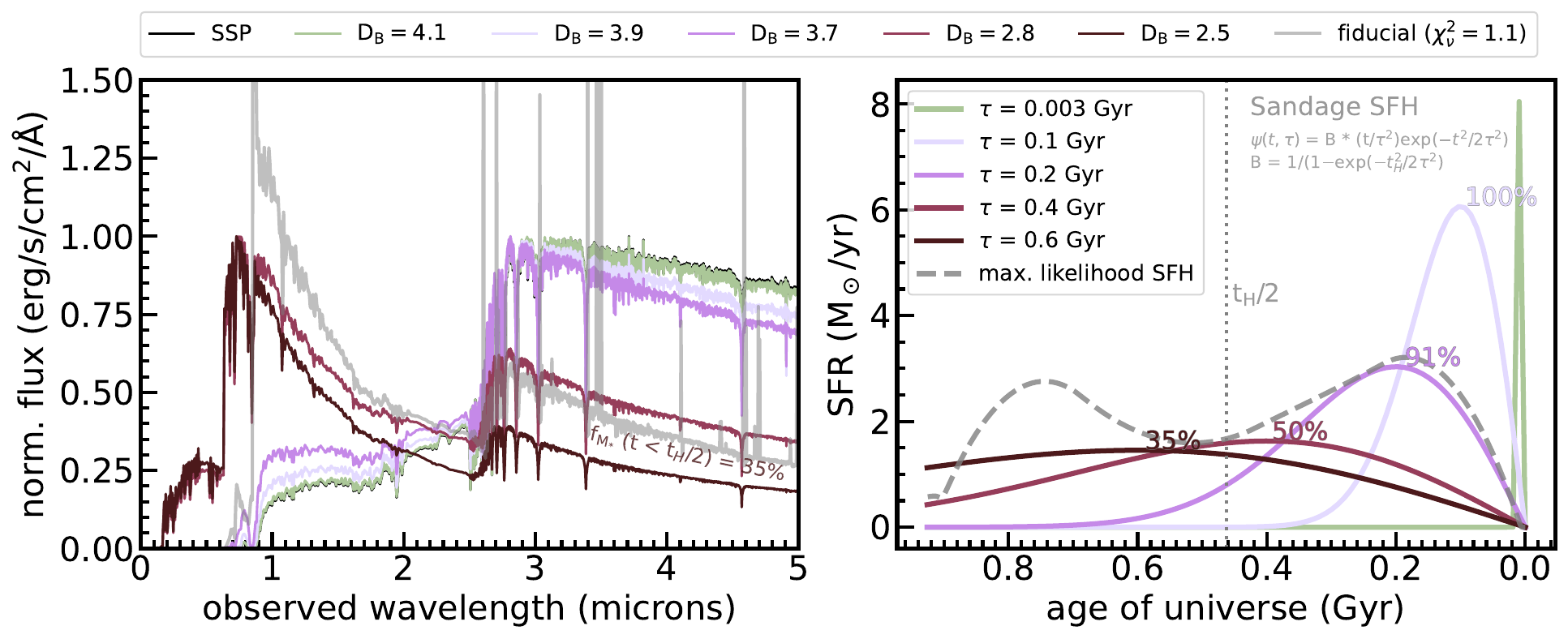}
\caption{Impact of the star formation history on a maximally-old stellar population at $z \approx 6$. The panels show stellar population models (left) created assuming the delayed-$\tau$ (Sandage) star formation histories of varying widths (right). The spectra are normalized to the peak flux of the instantaneous burst ($\tau = 3$ Myr) model. These models demonstrate how the strength of the Balmer break as the primary age indicator in the first billion years changes based on the SFH. Based on these, we define a maximally-old galaxy as one that formed more $> 50\%$ of its stars in half the Hubble time, $t_H(z)/2$. For comparison, we show the maximum-likelihood SFH and best-fit SED of \emph{The Sleeper} in gray. }\label{fig:max_old}
\end{figure}

To place the Balmer break strength of \emph{The Sleeper}  in the larger context of galaxy evolution in the first Gyr, we calculate the evolution of the break strengths of maximally-old stellar population models (Figure \ref{fig:max_old}). These models account for the dependence of $D_B$ on both the age and star formation history of the stellar population, assuming a fixed stellar metallicity of $10 \%$ (Sect. \ref{sec:mah}). Using \textsc{Fsps}, we generate a simple stellar population as old as the Universe at that redshift, $t = t_H (z)$, where $t_H$ is the Hubble time, and three models that parameterize our uncertainty about the onset of star formation as $t = t_H (z) - t_{delay}$, where $t_{delay} = 200,~ 300, ~\mathrm{and}~ 400 ~ \mathrm{Myr}$. All of these SSPs formed in an instantaneous burst, with $\tau = 0.003$ Gyr, using a delayed$-\tau$ model of the form: $\psi(t, \tau) = B  (t/\tau^2) e^{(-t^2/2\tau^2)}$, where $B = 1/(1-e^{(-t_H/2\tau^2)})$ is a normalization constant that determines the total mass of the stellar population. 

Because galaxies tend to have more extended star formation histories than an instantaneous burst due to several factors, including the local environment and feedback mechanisms, we also generate maximally-old stellar populations with $\tau = 200,~ 300, \mathrm{and}~ 400$ Myr. Based on these models, we define a maximally-old stellar population (at $z \gtrsim 5$) as one that formed at least 50\% of its stars in the first half of the Universe's history at the time of observation. We measure $D_B$ for each stellar population as a function of time and fit $D_B(z)$ with a polynomial to mitigate the impact of stellar absorption features in the continuum windows. The evolution of the Balmer break strength at $z = 5-10$ is then described by the following relation \edit{with coefficients given in Table \ref{tab:models}}: 
\begin{equation}
    D_B(z) = az^3 + bz^2 + cz + d
\end{equation}

\def\arraystretch{1.3}
\begin{table}
    \centering
    \caption{Coefficients for maximally-old stellar population models.}
    \begin{tabular}{l c c c c c }
    \hline
    \hline
Description & $t_{delay}$ [Myr] & \emph{a}  & \emph{b} & \emph{c} & \emph{d} \\
\hline

instantaneous burst & 0 &  $3.62 \times 10^{-3}$  & $-7.00 \times 10^{-2}$  & $0.298$ & $3.37$ \\
delayed instantaneous burst & 200 & $-3.18 \times 10^{-4}$  & $2.42 \times 10^{-2}$  & $-0.455$ & $5.16$ \\
delayed instantaneous burst & 300 & $-8.62 \times 10^{-4}$  & $3.94 \times 10^{-2}$  & $-0.596$ & $5.47$ \\
delayed instantaneous burst & 400 & $-3.68 \times 10^{-4}$  & $3.27 \times 10^{-2}$  & $-0.586$ & $5.43$ \\
extended burst $^\dagger$ & 0 &   $-1.01 \times 10^{-3}$  & $2.65 \times 10^{-2}$  & $-0.273$ & $3.29$ \\
    \hline
    \end{tabular}
    {\bf Note}:$^\dagger$ These assume $\tau(z)$ using the parameterization in Section \ref{sec:db_evolution} such that the fraction of stellar mass formed ($f_{M_\ast}$) at $t_H(z)/2$ is 50\%. This corresponds to $\tau = 400$ Myr at $z = 6$.
    \label{tab:models}
\end{table}

\subsection{Mass Assembly History and UV Luminosity Evolution}\label{sec:mah} 
The stellar mass of a given object is typically approximated as $M_\ast = M_{h} f_b \epsilon_\star$, assuming a cosmic baryon fraction of $f_b = 0.16$ \cite{planck2016}. $M_{h}$ is the dark halo mass for the galaxy assuming a fixed parameterization of the halo mass function and $\epsilon_\star$ is the efficiency at which baryons are converted into stars. While this approach is straightforward, it requires several assumptions about the selection function. This is often poorly characterized for the most massive galaxies at high redshift as they are rare and tend to cluster. This combined with the limited volumes of extragalactic surveys to makes a robust comparison between galaxy stellar masses and their likely dark matter haloes within those volumes extremely challenging.

The extreme value statistics (EVS)  approach (e.g., \cite{harrison_coles2012}) mitigates these issues by estimating the full probability density function of the most massive halo in a given survey volume, taking into account a physically-motivated distribution for $\epsilon_\star$ and a correction for Eddington bias. We calculate the mass assembly history of \emph{The Sleeper}  by integrating over the median star formation history in Figure \ref{fig:sfh}. \edit{This implicitly assumes that all stellar mass formed \emph{in situ} within a single progenitor halo, which may not hold in a hierarchical growth scenario e.g., \cite{cochrane2025}. However, given that galaxies of similar mass are predicted to form $\approx 90\%$ of their stellar mass in-situ at $z \approx 6$ \cite{rg2016}, this assumption holds.} We compare this assembly history to the halo mass function at high redshift using the \textsc{Evstats} package \cite{lovell2023} assuming the survey area of 29.1 sq arcmin covered by the three CANUCS/Technicolor flanking fields. 

Using this mass assembly history, we predict the UV luminosity evolution of \emph{The Sleeper} from $z \approx 6 - 20$ using models that match its observed properties and those of JADES-GS-14-0. At each time step, we use \textsc{Bagpipes} to generate a stellar population with the stellar mass and star formation rate of \emph{The Sleeper} at that redshift, assuming the same parameterization as outlined in Sect. \ref{sec:dense_basis}. We fix the metallicity at $10\%$ solar based on the mass-metallicity relation at $z = 5 - 12$ from the FIRE-2 simulation \cite{marszewski2024}, assuming that the gas-phase and stellar metallicities are coupled and that there is negligible evolution in the mass-metallicity relation at $z > 12$. \edit{To match the properties and fitting assumptions used for JADES-GS-z14-0 \cite{carniani2025}}, we use a fixed ionization parameter of $log(U) = -2.39$ and an SMC dust law. We allow dust to evolve monotonically from A$_\mathrm{V} \approx 0.1$ at $z \sim 15$ to A$_\mathrm{V} \approx 0.4$ at $z \sim 6$. This captures the uncertainty in dust measurements (Sect. \ref{sec:dust}) while accounting for the data preference for low but non-zero dust in both populations. We then estimate the UV luminosity as a function of redshift using the model photometry as described in Sect. \ref{sec:empirical}. 

\begin{figure}[t!]%
\centering
\includegraphics[trim={0.2cm 0.1cm 0.1cm 0.1cm}, clip, scale=0.265]{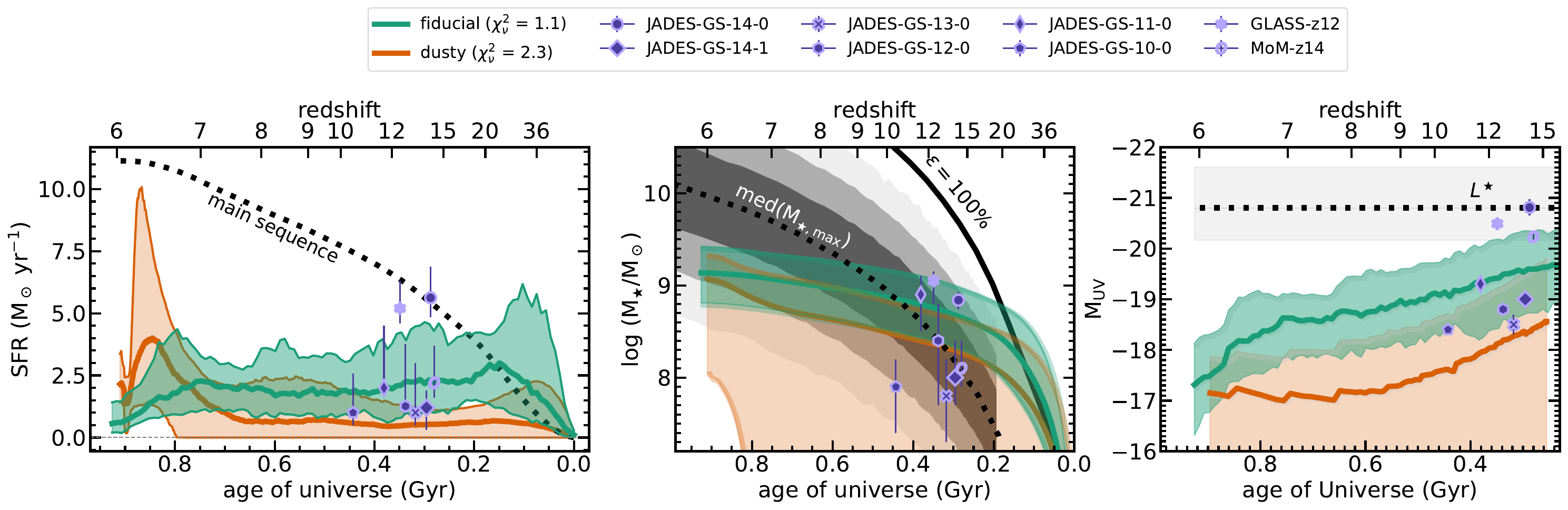}
\caption{Star formation histories, stellar mass assembly histories, and UV luminosity evolution of \emph{The Sleeper}  based on the fiducial and dusty solutions. Together, these two solutions bracket the range of possible assembly histories for the luminous $ z > 10$ population if they do not undergo mergers following their initial period of bursty star formation. The star formation rates of the literature sample are averaged over $30 - 50$ Myr timescales. The gray shaded regions in the middle panel show the $1\sigma, 2\sigma$, and $3\sigma$ confidence intervals of the probability density function of the most massive halo in the volume probed by a $\approx 29$ sq. arcmin survey and the black line is the theoretical maximum.}\label{fig:sfhs_comparison}
\end{figure} 

\subsection{Emission Line Measurements}\label{sec:em_lines}
Following \cite{papovich2001}, we estimate emission line equivalent widths by assuming a roughly flat continuum around the line
\begin{equation}
W_0 \simeq \bigg{(}\frac{F_{\lambda}}{F^{c}_{\lambda}} - 1\bigg{)} \times \frac{\Delta \lambda}{1 + z} , 
\end{equation}
with the errors propagated accordingly 
\begin{equation}
\frac{\sigma_{W_0}}{W_0} = \sqrt{\bigg{(}\frac{\sigma_{{F_{\lambda}}}}{F_{\lambda}}\bigg{)}^2 + \bigg{(}\frac{\sigma_{{F^{c}_{\lambda}}}}{F^{c}_{\lambda}} \bigg{)}^2}
\end{equation}
where $F_{\lambda}$ is the flux density measured through the medium band filter capturing the emission line and $F^{c}_{\lambda}$ is the continuum flux density. Assuming the best-fit SED from the fiducial model (Sect. \ref{sec:dense_basis}), we identify F356W and F480M as the continuum filters for \OIII + \Hbeta and \Halpha + \NII, respectively, and use F335M and F460M for their emission line flux measurements. We do not account for the varying filter response functions. Because the transmission curves are nearly flat, we expect a $< 10\%$ change in flux \cite{lorenz2025}. The measured equivalent widths are reported in Table \ref{tab:summary}.

\subsection{Comparison to Cosmological Hydrodynamical Simulations}
\label{sec:sims}
We compare the properties of \emph{The Sleeper} to those of galaxies from three large-volume cosmological simulations specifically targeted at tracing the evolution of galaxies in the Epoch of Reionization (EoR). The First Light And Reionisation Epoch Simulations (\textsc{Flares} \cite{lovell2021}) uses the \textsc{Eagle} \cite{schaye2015} model to conduct zoom simulations of overdensities in a 3.2 comoving Gpc$^3$ volume. This approach makes \textsc{Flares} ideal for studying rare populations in the early Universe with sufficient resolution to explore the details of the physical conditions governing star formation and quenching. We also use spectra from the \textsc{Sphinx$^{20}$} public data release \cite{katz2023}, simulated over a $20^3$ Mpc$^3$ comoving volume with radiative transfer and high enough resolution on the scale of small molecular clouds ($\approx 10$ pc), enabling detailed modeling of the dynamics and phase structure of the ISM. This is crucial for dust and emission line predictions. Finally, we use spectra from the \textsc{thesan} simulation \cite{kannan2022_spectra}, which builds on the IllustrisTNG model \cite{pillepich2018} using on-the-fly radiative hydrodynamics in a box of length $95.5$ Mpc. Notably, both \textsc{thesan} and \textsc{Flares} include empirical prescriptions for black hole formation and feedback, which are especially important for quenching \edit{massive galaxies} at high redshift (e.g., \cite{park2024}). 

\begin{figure}[t]%
\centering
\includegraphics[trim={2.5cm 2cm 4.5cm 2cm}, scale = 0.175]{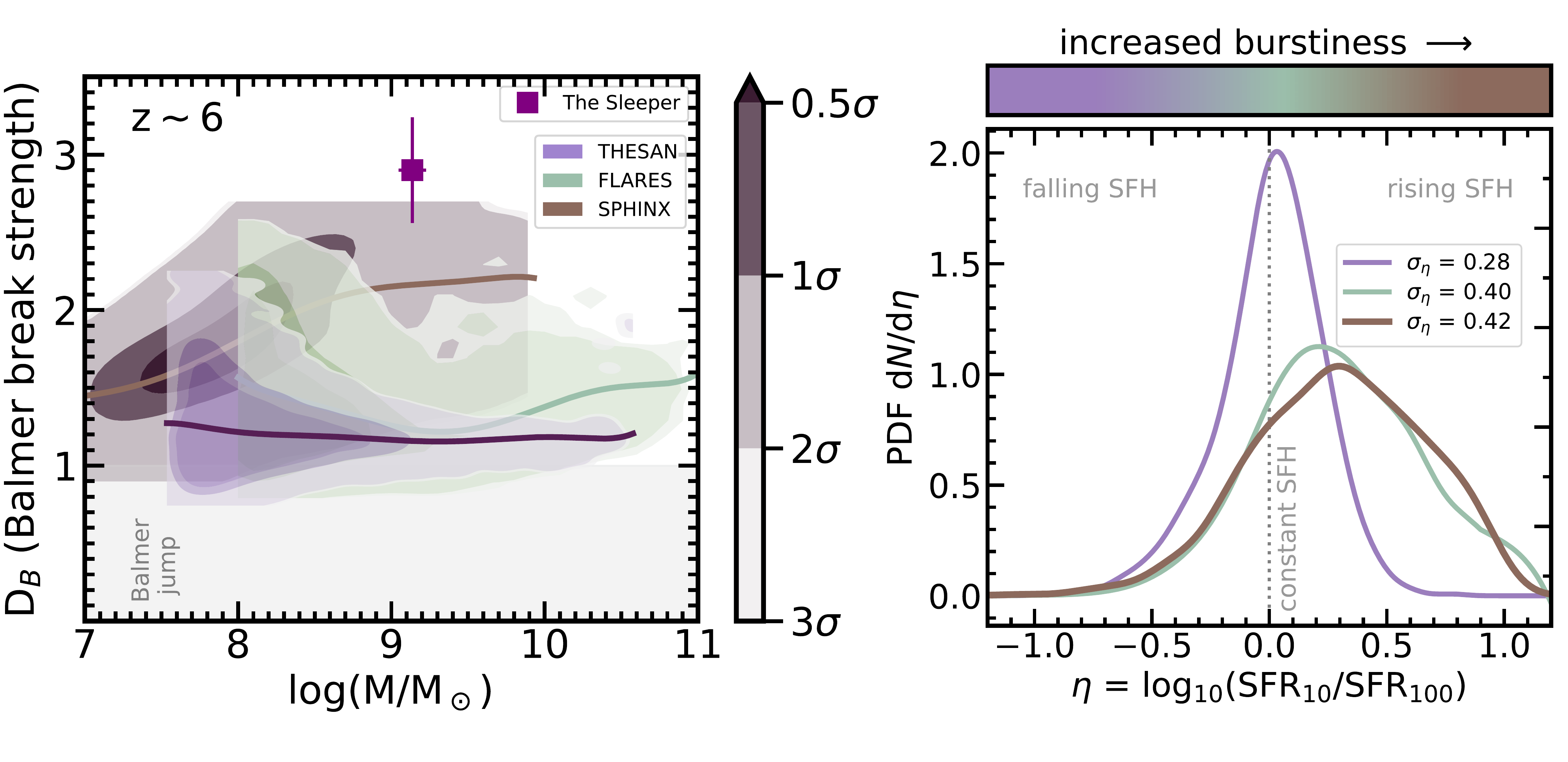}
\caption{Relationship between burstiness and Balmer break strength in cosmological simulations. Left: The contours show the $0.5 - 3 \sigma$ confidence intervals (dark to light) of the intrinsic Balmer break strength distributions of galaxies in \textsc{Sphinx$^{20}$}, \textsc{thesan}, and \textsc{Flares} at $z \sim 6$. The kernel density estimates are truncated at the respective mass limit of each simulation. Solid lines show the median break strength as a function of stellar mass. \emph{The Sleeper}  is extremely rare in these simulations, having a Balmer break strength that is a $\sim 2-4 \sigma$ outlier from that of the typical galaxy at $z \sim 6$. Right: We explore the relationship between burstiness and Balmer break strength using the width of the burst intensity, $\eta$. For reference, \emph{The Sleeper} has $\eta = -0.25_{-0.15}^{+0.04}$, due to its rapid decline in star formation rate over the past 100 Myr. The simulations with the largest scatter in burst intensity produce the strongest breaks.  }\label{fig:sims}
\end{figure}

In each simulation, we evaluate the probability of finding galaxies with the break strength of \emph{The Sleeper}. The break strength is measured from the intrinsic spectra of galaxies at $z \sim 6$ using the same windows and methodology in Sect. \ref{sec:db_evolution}. We show the distribution of Balmer break strength as a function of stellar mass for each simulation in Figure \ref{fig:sims}. To account for the measurement uncertainty, we perturb these distributions by $\sigma_\mathrm{D_B} = 0.34$ and $\sigma_\mathrm{M_\ast} = 0.09$. We then estimate the probability density function of each $D_B - M_\ast$ distribution using a kernel density estimate. We then evaluate the probability of finding a galaxy in each distribution with $D_B \geq 2.9$ using the Mahalanobis metric, \edit{which permits multivariate, covariance-weighted distances}. We find that \emph{The Sleeper} is extremely rare in all three simulations, being a $2.15\sigma,~ 3.10\sigma,$ and $4.24\sigma$ outlier in \textsc{Sphinx$^{20}$},  \textsc{Flares}, and \textsc{thesan}, respectively. We note that because the kernel density estimates in each simulation are non-Gaussian, the Mahalanobis distance is likely an underestimate. 

 The CANUCS/Technicolor effective survey area ($21.72$ sq. arcmin \cite{desprez2024}) implies an observed number density of $n = 0.34_{-0.79}^{+1.69} \times 10^{-4}~ \mathrm{Mpc^{-3}}$ at $z = 5 - 6$, based on the photometric selection of two galaxies meeting the criteria: log$(M_\ast/M_\odot) > 9$ and $D_B \geq 2.69$. There are no such galaxies in any of the three simulations, although \textsc{Sphinx$^{20}$} produces the strongest Balmer breaks. This sets an upper limit of $n < 1.25 \times 10^{-4}~ \mathrm{ Mpc^{-3}}$, $n < 3.13 \times 10^{-6} ~\mathrm{Mpc^{-3}}$, and $n < 1.15 \times 10^{-6}~ \mathrm{ Mpc^{-3}}$ on the number density in \textsc{Sphinx$^{20}$},  \textsc{Flares}, and  \textsc{thesan}, respectively. The relatively large volumes of \textsc{thesan} and \textsc{Flares}, as well as the selection strategy of the latter suggest that the rarity of such relics in these simulations is not due to cosmic variance, but rather due to the assumptions made regarding the underlying physics governing galaxy evolution at high redshift. 
In simulations, periods of enhanced star formation are naturally followed by periods of suppressed star formation due to the opposing effects of stellar feedback and gravity \cite{fg2018}. The more massive the galaxy, the longer it is able to sustain equilibrium between these two processes, typically reaching steady star formation at $\approx 10^9 M_\odot$ \cite{gelli2025}. Therefore, at fixed stellar mass, the galaxies with the largest burst intensities will experience longer periods of suppressed star formation. In other words, we expect the simulations with the highest burstiness to have the strongest Balmer breaks. 

We quantify burstiness using an indicator that compares a galaxy's recent and past star formation,  $\eta = \mathrm{log_{10}(SFR_{10}/SFR_{100})}$ (e.g., \cite{broussard2019}). We estimate the width, $\sigma_\eta$, of a skewed Gaussian (\edit{classical}) fit to this distribution in each simulation at $z \sim 6$. As predicted, \textsc{Sphinx$^{20}$} has the largest scatter in burst intensity, $\sigma_\eta = 0.42$ (Figure \ref{fig:sims}). This is in spite of its selection function, SFR$_{10}$ $\geq 0.3~ \mathrm{M_\odot~ yr^{-1}}$, which is skewed towards galaxies that have undergone a recent burst in star formation ($\eta > 0$). In comparison, although \textsc{thesan} has a higher fraction of galaxies with a recent decline in their star formation rates ($\eta < 0$) due to its selection function of SFR$_{10} \geq 0.1~ \mathrm{M_\odot~ yr^{-1}}$, it has the smallest burst intensity scatter, $\sigma_\eta = 0.28$, in line with the dearth of strong Balmer breaks at log$(M_\ast/M_\odot) > 9$. \textsc{Flares} has a scatter comparable to that of \textsc{Sphinx$^{20}$}, indicative of increased burstiness in these simulations, as other studies have found \cite{kokorev2025}. However, similar to \textsc{thesan}, it lacks galaxies with strong Balmer breaks at log$(M_\ast/M_\odot) > 9$. Importantly, their distribution in the $D_B - M_\ast$ plane is remarkably similar, with a relatively flat median Balmer break strength as a function of stellar mass and the fraction of strong breaks being skewed towards low stellar masses (log$(M_\ast/M_\odot) < 9$). \edit{In \textsc{thesan}, the interstellar medium is modelled as a two-phase structure of gas where cold clumps are embedded in a smooth, hot phase produced by supernova explosions \cite{springel2003}. This model tends to produce relatively smooth and non-bursty star formation histories, when compared to models that aim to simulate the multi-phase nature of the ISM (Marinacci et al. 2019). \textsc{Flares} does not model the cold gas phase \cite{schaye2015}.} This suggests that resolving the multiphase ISM may play an important role in understanding quenching in the EoR. 

\subsection{The Environment of \emph{The Sleeper} }\label{sec:overdensity}
\emph{The Sleeper} resides in a flanking field of the Abell 370 galaxy cluster and is separated by $\approx0.5"$ from a $\log(M_\star/M_\odot) = 9.55 \pm 0.35$ solar mass companion at $z = 6.00 \pm 0.03$, which corresponds to a projected distance of $\approx 3$ kpc at this redshift. The high photometric sampling strongly constrains their redshifts, which overlap within the $1\sigma$ uncertainty. Consequently, these galaxies are very likely physically associated and not merely a chance projection. The companion also shows a detected Balmer break in the form of an excess in the F250M and  F300M medium bands. We also observe a significant number of Balmer break galaxies at $z=5.87 - 6.13$.  This overdensity contains 37 members, including \emph{The Sleeper}.  Of these, 75\% are in close pairs and n-tuples, at a projected distance of $R < 5$ kpc from at least one other companion. These properties place \emph{The Sleeper} within a remarkably evolved overdensity at $z\sim6$ e.g., \cite{witten2025_ovd, morishita_ovd}. 

\backmatter

\bmhead{Supplementary information}

\bmhead{Acknowledgements}
The authors are grateful to the 20,000 scientists, engineers, and administrators across Canada, the US, and Europe who made the James Webb Space Telescope a reality and birthed an era of unprecedented discovery in astronomy. We thank Chris Lovell and Harley Katz for sharing data and code that were used in the comparison with the \textsc{Flares} and \textsc{Sphinx$^{20}$} simulations, respectively. We also thank David Setton and Adarsh Kuruvanthodi for providing data used in this analysis. Finally, we are immensely grateful to Justin Cole for useful discussions about non-parametric SED fitting with Bagpipes and John Weaver for conversations that helped frame the narrative about \emph{The Sleeper}. This research made use of the {\tt pathfinder} LLM-enabled literature search framework \cite{iyer2024}.

\section*{Declarations}

\begin{itemize}
\item \textbf{Funding} Antwi-Danso is supported by a Banting Fellowship from the Natural Sciences and Engineering Research Council of Canada (NSERC). The Dunlap Institute is funded through an endowment established by the David Dunlap family and the University of Toronto. Marchesini and Robbins gratefully acknowledge funding from JWST-GO-3362, provided through a grant from the STScI under NASA contract NAS5-03127. Asada is supported by a Research Fellowship for Young Scientists from the Japan Society of the Promotion of Science (JSPS) and by JSPS KAKENHI Grant Number 23H00131. \sur{Brada\v{c}} acknowledges support from the ERC Grant FIRST- LIGHT and from the Slovenian national research agency ARRS through grants N1-0238, P1-0188 and the program HST-GO-16667, provided through a grant from the STScI under NASA contract NAS5-26555. This research used the Canadian Advanced Network For Astronomy Research (CANFAR) operated in partnership by the Canadian Astronomy Data Centre and The Digital Research Alliance of Canada with support from the National Research Council of Canada the Canadian Space Agency, CANARIE, and the Canadian Foundation for Innovation. Kannan acknowledges support of the Natural Sciences and Engineering Research Council of Canada (NSERC) through a Discovery Grant anda Discovery Launch Supplement (funding reference numbers RGPIN-2024-06222 and  DGECR-2024-00144) and York University’s Global Research Excellence Initiative.

\item \textbf{Data availability} 
The imaging, photometric catalogs, and stellar population properties from the CANUCS and Technicolor surveys are publicly available and accessible via the Mikulski Archive for Space Telescopes (\href{https://archive.stsci.edu/hlsp/canucs}{MAST}). The spectra for galaxies in the SPHINX cosmological simulation are also accessible via their repository on \href{https://github.com/HarleyKatz/SPHINX-20-data}{Github}. 

\item \textbf{Code availability} 
We use the following publicly-available codes for the analysis presented in this work: \href{https://github.com/kartheikiyer/dense_basis}{\tt DENSE BASIS}, \href{https://github.com/ACCarnall/bagpipes}{\textsc{Bagpipes}}, \href{https://github.com/gbrammer/eazy-py}{\tt eazy-py}, \href{https://github.com/gbrammer/grizli}{\tt Grizli}, and \href{https://github.com/christopherlovell/evstats}{\textsc{evstats}}. 

\item \textbf{Author contribution}
Antwi-Danso led the analysis and writing of this paper, with key contributions from Robbins, Muzzin, Marchesini, Asada, Sawicki, Speagle, Whitaker, and Papovich. Willott and Muzzin desiged the CANUCS and Technicolor surveys on which this work is based. Sarrouh, Asada, Martis, Noirot, and Desprez led the image processing and data reduction. Iyer and Martis generated the spectrophotometric catalogs from which the target was identified. Kannan and Speagle contributed to the comparison with cosmological hydrodynamical simulations and statistical tests. All authors reviewed the manuscript and provided invaluable assistance in the data analysis and interpretation. 
\end{itemize}

\subsection*{Competing interests Statement} The authors declare no competing interests.

\bibliography{sn-bibliography}

\begin{thebibliography}{100}
\expandafter\ifx\csname url\endcsname\relax
  \def\url#1{\burl{#1}}\fi
\expandafter\ifx\csname urlprefix\endcsname\relax\def\urlprefix{URL }\fi
\providecommand{\bibinfo}[2]{#2}
\providecommand{\eprint}[2][]{\url{#2}}
\providecommand{\doi}[1]{\url{https://doi.org/#1}}
\bibcommenthead

\bibitem{donnan2024}
\bibinfo{author}{{Donnan}, C.~T.} \emph{et~al.}
\newblock \bibinfo{title}{{JWST PRIMER: a new multifield determination of the evolving galaxy UV luminosity function at redshifts $z \approx 9 - 15$}}.
\newblock \emph{\bibinfo{journal}{\mnras}} \textbf{\bibinfo{volume}{533}}, \bibinfo{pages}{3222--3237} (\bibinfo{year}{2024}).

\bibitem{chmerynska2024}
\bibinfo{author}{{Chemerynska}, I.} \emph{et~al.}
\newblock \bibinfo{title}{{JWST UNCOVER: the overabundance of ultraviolet-luminous galaxies at z > 9}}.
\newblock \emph{\bibinfo{journal}{\mnras}} \textbf{\bibinfo{volume}{531}}, \bibinfo{pages}{2615--2625} (\bibinfo{year}{2024}).

\bibitem{sun2023}
\bibinfo{author}{{Sun}, G.} \emph{et~al.}
\newblock \bibinfo{title}{{Bursty Star Formation Naturally Explains the Abundance of Bright Galaxies at Cosmic Dawn}}.
\newblock \emph{\bibinfo{journal}{\apjl}} \textbf{\bibinfo{volume}{955}}, \bibinfo{pages}{L35} (\bibinfo{year}{2023}).

\bibitem{cole2025}
\bibinfo{author}{{Cole}, J.~W.} \emph{et~al.}
\newblock \bibinfo{title}{{CEERS: Increasing Scatter along the Star-forming Main Sequence Indicates Early Galaxies Form in Bursts}}.
\newblock \emph{\bibinfo{journal}{\apj}} \textbf{\bibinfo{volume}{979}}, \bibinfo{pages}{193} (\bibinfo{year}{2025}).

\bibitem{dome2024}
\bibinfo{author}{{Dome}, T.} \emph{et~al.}
\newblock \bibinfo{title}{{Mini-quenching of z = 4-8 galaxies by bursty star formation}}.
\newblock \emph{\bibinfo{journal}{\mnras}} \textbf{\bibinfo{volume}{527}}, \bibinfo{pages}{2139--2151} (\bibinfo{year}{2024}).

\bibitem{gelli2025}
\bibinfo{author}{{Gelli}, V.} \emph{et~al.}
\newblock \bibinfo{title}{{Temporarily Quiescent Galaxies at Cosmic Dawn: Probing Bursty Star Formation}}.
\newblock \emph{\bibinfo{journal}{\apj}} \textbf{\bibinfo{volume}{985}}, \bibinfo{pages}{126} (\bibinfo{year}{2025}).

\bibitem{carniani2024}
\bibinfo{author}{{Carniani}, S.} \emph{et~al.}
\newblock \bibinfo{title}{{Spectroscopic confirmation of two luminous galaxies at a redshift of 14}}.
\newblock \emph{\bibinfo{journal}{\nat}} \textbf{\bibinfo{volume}{633}}, \bibinfo{pages}{318--322} (\bibinfo{year}{2024}).

\bibitem{carniani2025}
\bibinfo{author}{{Carniani}, S.} \emph{et~al.}
\newblock \bibinfo{title}{{The eventful life of a luminous galaxy at z = 14: metal enrichment, feedback, and low gas fraction?}}
\newblock \emph{\bibinfo{journal}{\aap}} \textbf{\bibinfo{volume}{696}}, \bibinfo{pages}{A87} (\bibinfo{year}{2025}).

\bibitem{robertson2023}
\bibinfo{author}{{Robertson}, B.~E.} \emph{et~al.}
\newblock \bibinfo{title}{{Identification and properties of intense star-forming galaxies at redshifts z > 10}}.
\newblock \emph{\bibinfo{journal}{Nature Astronomy}} \textbf{\bibinfo{volume}{7}}, \bibinfo{pages}{611--621} (\bibinfo{year}{2023}).

\bibitem{naidu2025}
\bibinfo{author}{{Naidu}, R.~P.} \emph{et~al.}
\newblock \bibinfo{title}{{A Cosmic Miracle: A Remarkably Luminous Galaxy at $z_{\rm{spec}}=14.44$ Confirmed with JWST}}.
\newblock \emph{\bibinfo{journal}{arXiv e-prints}} \bibinfo{pages}{arXiv:2505.11263} (\bibinfo{year}{2025}).

\bibitem{castellano2024}
\bibinfo{author}{{Castellano}, M.} \emph{et~al.}
\newblock \bibinfo{title}{{JWST NIRSpec Spectroscopy of the Remarkable Bright Galaxy GHZ2/GLASS-z12 at Redshift 12.34}}.
\newblock \emph{\bibinfo{journal}{\apj}} \textbf{\bibinfo{volume}{972}}, \bibinfo{pages}{143} (\bibinfo{year}{2024}).

\bibitem{looser2024}
\bibinfo{author}{{Looser}, T.~J.} \emph{et~al.}
\newblock \bibinfo{title}{{A recently quenched galaxy 700 million years after the Big Bang}}.
\newblock \emph{\bibinfo{journal}{\nat}} \textbf{\bibinfo{volume}{629}}, \bibinfo{pages}{53--57} (\bibinfo{year}{2024}).

\bibitem{witten2025}
\bibinfo{author}{{Witten}, C.} \emph{et~al.}
\newblock \bibinfo{title}{{Rising from the ashes: evidence of old stellar populations and rejuvenation events in the very early Universe}}.
\newblock \emph{\bibinfo{journal}{\mnras}} \textbf{\bibinfo{volume}{537}}, \bibinfo{pages}{112--126} (\bibinfo{year}{2025}).

\bibitem{vikaeus2024}
\bibinfo{author}{{Vikaeus}, A.} \emph{et~al.}
\newblock \bibinfo{title}{{To be, or not to be: Balmer breaks in high-z galaxies with JWST}}.
\newblock \emph{\bibinfo{journal}{\mnras}} \textbf{\bibinfo{volume}{529}}, \bibinfo{pages}{1299--1307} (\bibinfo{year}{2024}).

\bibitem{strait2023}
\bibinfo{author}{{Strait}, V.} \emph{et~al.}
\newblock \bibinfo{title}{{An Extremely Compact, Low-mass Galaxy on its Way to Quiescence at z = 5.2}}.
\newblock \emph{\bibinfo{journal}{\apjl}} \textbf{\bibinfo{volume}{949}}, \bibinfo{pages}{L23} (\bibinfo{year}{2023}).

\bibitem{lu2025}
\bibinfo{author}{{Lu}, S.} \emph{et~al.}
\newblock \bibinfo{title}{{A comparison of pre-existing {\ensuremath{\Lambda}}CDM predictions with the abundance of JWST galaxies at high redshift}}.
\newblock \emph{\bibinfo{journal}{\mnras}} \textbf{\bibinfo{volume}{536}}, \bibinfo{pages}{1018--1034} (\bibinfo{year}{2025}).

\bibitem{harikane2023}
\bibinfo{author}{{Harikane}, Y.} \emph{et~al.}
\newblock \bibinfo{title}{{A Comprehensive Study of Galaxies at z 9-16 Found in the Early JWST Data: Ultraviolet Luminosity Functions and Cosmic Star Formation History at the Pre-reionization Epoch}}.
\newblock \emph{\bibinfo{journal}{\apjs}} \textbf{\bibinfo{volume}{265}}, \bibinfo{pages}{5} (\bibinfo{year}{2023}).

\bibitem{weibel2025}
\bibinfo{author}{{Weibel}, A.} \emph{et~al.}
\newblock \bibinfo{title}{{RUBIES Reveals a Massive Quiescent Galaxy at z = 7.3}}.
\newblock \emph{\bibinfo{journal}{\apj}} \textbf{\bibinfo{volume}{983}}, \bibinfo{pages}{11} (\bibinfo{year}{2025}).

\bibitem{paquereau2025}
\bibinfo{author}{{Paquereau}, L.} \emph{et~al.}
\newblock \bibinfo{title}{{Tracing the galaxy-halo connection with galaxy clustering in COSMOS-Web from z = 0.1 to z \raisebox{-0.5ex}\textasciitilde 12}}.
\newblock \emph{\bibinfo{journal}{arXiv e-prints}} \bibinfo{pages}{arXiv:2501.11674} (\bibinfo{year}{2025}).

\bibitem{wyithe2014}
\bibinfo{author}{{Wyithe}, J. S.~B.}, \bibinfo{author}{{Loeb}, A.} \& \bibinfo{author}{{Oesch}, P.~A.}
\newblock \bibinfo{title}{{A predicted new population of UV-faint galaxies at z {\ensuremath{\gtrsim}} 4}}.
\newblock \emph{\bibinfo{journal}{\mnras}} \textbf{\bibinfo{volume}{439}}, \bibinfo{pages}{1326--1336} (\bibinfo{year}{2014}).

\bibitem{willott2024}
\bibinfo{author}{{Willott}, C.~J.} \emph{et~al.}
\newblock \bibinfo{title}{{A Steep Decline in the Galaxy Space Density beyond Redshift 9 in the CANUCS UV Luminosity Function}}.
\newblock \emph{\bibinfo{journal}{\apj}} \textbf{\bibinfo{volume}{966}}, \bibinfo{pages}{74} (\bibinfo{year}{2024}).

\bibitem{willott2022}
\bibinfo{author}{{Willott}, C.~J.} \emph{et~al.}
\newblock \bibinfo{title}{{The Near-infrared Imager and Slitless Spectrograph for the James Webb Space Telescope. II. Wide Field Slitless Spectroscopy}}.
\newblock \emph{\bibinfo{journal}{\pasp}} \textbf{\bibinfo{volume}{134}}, \bibinfo{pages}{025002} (\bibinfo{year}{2022}).

\bibitem{lotz2017}
\bibinfo{author}{{Lotz}, J.~M.} \emph{et~al.}
\newblock \bibinfo{title}{{The Frontier Fields: Survey Design and Initial Results}}.
\newblock \emph{\bibinfo{journal}{\apj}} \textbf{\bibinfo{volume}{837}}, \bibinfo{pages}{97} (\bibinfo{year}{2017}).

\bibitem{iyer2017}
\bibinfo{author}{{Iyer}, K.} \& \bibinfo{author}{{Gawiser}, E.}
\newblock \bibinfo{title}{{Reconstruction of Galaxy Star Formation Histories through SED Fitting:The Dense Basis Approach}}.
\newblock \emph{\bibinfo{journal}{\apj}} \textbf{\bibinfo{volume}{838}}, \bibinfo{pages}{127} (\bibinfo{year}{2017}).

\bibitem{iyer2019}
\bibinfo{author}{{Iyer}, K.~G.} \emph{et~al.}
\newblock \bibinfo{title}{{Nonparametric Star Formation History Reconstruction with Gaussian Processes. I. Counting Major Episodes of Star Formation}}.
\newblock \emph{\bibinfo{journal}{\apj}} \textbf{\bibinfo{volume}{879}}, \bibinfo{pages}{116} (\bibinfo{year}{2019}).

\bibitem{rb2024}
\bibinfo{author}{{Roberts-Borsani}, G.} \emph{et~al.}
\newblock \bibinfo{title}{{Between the Extremes: A JWST Spectroscopic Benchmark for High-redshift Galaxies Using {\ensuremath{\sim}}500 Confirmed Sources at z {\ensuremath{\geq}} 5}}.
\newblock \emph{\bibinfo{journal}{\apj}} \textbf{\bibinfo{volume}{976}}, \bibinfo{pages}{193} (\bibinfo{year}{2024}).

\bibitem{gonzelez-delgado1999}
\bibinfo{author}{{Gonz{\'a}lez Delgado}, R.~M.}, \bibinfo{author}{{Leitherer}, C.} \& \bibinfo{author}{{Heckman}, T.~M.}
\newblock \bibinfo{title}{{Synthetic Spectra of H Balmer and HE I Absorption Lines. II. Evolutionary Synthesis Models for Starburst and Poststarburst Galaxies}}.
\newblock \emph{\bibinfo{journal}{\apjs}} \textbf{\bibinfo{volume}{125}}, \bibinfo{pages}{489--509} (\bibinfo{year}{1999}).

\bibitem{endsley2024}
\bibinfo{author}{{Endsley}, R.} \emph{et~al.}
\newblock \bibinfo{title}{{The star-forming and ionizing properties of dwarf z 6-9 galaxies in JADES: insights on bursty star formation and ionized bubble growth}}.
\newblock \emph{\bibinfo{journal}{\mnras}} \textbf{\bibinfo{volume}{533}}, \bibinfo{pages}{1111--1142} (\bibinfo{year}{2024}).

\bibitem{covelo-paz2025}
\bibinfo{author}{{Covelo-Paz}, A.} \emph{et~al.}
\newblock \bibinfo{title}{{A systematic search for dormant galaxies at z\raisebox{-0.5ex}\textasciitilde5-7 from the JWST NIRSpec archive}}.
\newblock \emph{\bibinfo{journal}{arXiv e-prints}} \bibinfo{pages}{arXiv:2506.22540} (\bibinfo{year}{2025}).

\bibitem{katz2023}
\bibinfo{author}{{Katz}, H.} \emph{et~al.}
\newblock \bibinfo{title}{{The SPHINX Public Data Release: Forward Modelling High-Redshift JWST Observations with Cosmological Radiation Hydrodynamics Simulations}}.
\newblock \emph{\bibinfo{journal}{The Open Journal of Astrophysics}} \textbf{\bibinfo{volume}{6}}, \bibinfo{pages}{44} (\bibinfo{year}{2023}).

\bibitem{lovell2021}
\bibinfo{author}{{Lovell}, C.~C.} \emph{et~al.}
\newblock \bibinfo{title}{{First Light And Reionization Epoch Simulations (FLARES) - I. Environmental dependence of high-redshift galaxy evolution}}.
\newblock \emph{\bibinfo{journal}{\mnras}} \textbf{\bibinfo{volume}{500}}, \bibinfo{pages}{2127--2145} (\bibinfo{year}{2021}).

\bibitem{kannan2022}
\bibinfo{author}{{Kannan}, R.} \emph{et~al.}
\newblock \bibinfo{title}{{Introducing the THESAN project: radiation-magnetohydrodynamic simulations of the epoch of reionization}}.
\newblock \emph{\bibinfo{journal}{\mnras}} \textbf{\bibinfo{volume}{511}}, \bibinfo{pages}{4005--4030} (\bibinfo{year}{2022}).

\bibitem{chiang2017}
\bibinfo{author}{{Chiang}, Y.-K.}, \bibinfo{author}{{Overzier}, R.~A.}, \bibinfo{author}{{Gebhardt}, K.} \& \bibinfo{author}{{Henriques}, B.}
\newblock \bibinfo{title}{{Galaxy Protoclusters as Drivers of Cosmic Star Formation History in the First 2 Gyr}}.
\newblock \emph{\bibinfo{journal}{\apjl}} \textbf{\bibinfo{volume}{844}}, \bibinfo{pages}{L23} (\bibinfo{year}{2017}).

\bibitem{desprez2024}
\bibinfo{author}{{Desprez}, G.} \emph{et~al.}
\newblock \bibinfo{title}{{{\ensuremath{\Lambda}}CDM not dead yet: massive high-z Balmer break galaxies are less common than previously reported}}.
\newblock \emph{\bibinfo{journal}{\mnras}} \textbf{\bibinfo{volume}{530}}, \bibinfo{pages}{2935--2952} (\bibinfo{year}{2024}).

\bibitem{suess2024}
\bibinfo{author}{{Suess}, K.~A.} \emph{et~al.}
\newblock \bibinfo{title}{{Medium Bands, Mega Science: A JWST/NIRCam Medium-band Imaging Survey of A2744}}.
\newblock \emph{\bibinfo{journal}{\apj}} \textbf{\bibinfo{volume}{976}}, \bibinfo{pages}{101} (\bibinfo{year}{2024}).

\bibitem{willott2025}
\bibinfo{author}{{Willott}, C.~J.} \emph{et~al.}
\newblock \bibinfo{title}{{In Search of the First Stars: An Ultra-compact and Very-low-metallicity Ly{\ensuremath{\alpha}} Emitter Deep within the Epoch of Reionization}}.
\newblock \emph{\bibinfo{journal}{\apj}} \textbf{\bibinfo{volume}{988}}, \bibinfo{pages}{26} (\bibinfo{year}{2025}).

\bibitem{hsiao2025}
\bibinfo{author}{{Hsiao}, T. Y.-Y.} \emph{et~al.}
\newblock \bibinfo{title}{{SAPPHIRES: Extremely Metal-Poor Galaxy Candidates with $12+{\rm log(O/H)}<7.0$ at $z\sim5-7$ from Deep JWST/NIRCam Grism Observations}}.
\newblock \emph{\bibinfo{journal}{arXiv e-prints}} \bibinfo{pages}{arXiv:2505.03873} (\bibinfo{year}{2025}).

\bibitem{algera2023}
\bibinfo{author}{{Algera}, H. S.~B.} \emph{et~al.}
\newblock \bibinfo{title}{{The ALMA REBELS survey: the dust-obscured cosmic star formation rate density at redshift 7}}.
\newblock \emph{\bibinfo{journal}{\mnras}} \textbf{\bibinfo{volume}{518}}, \bibinfo{pages}{6142--6157} (\bibinfo{year}{2023}).

\bibitem{md2023}
\bibinfo{author}{{Mauerhofer}, V.} \& \bibinfo{author}{{Dayal}, P.}
\newblock \bibinfo{title}{{The dust enrichment of early galaxies in the JWST and ALMA era}}.
\newblock \emph{\bibinfo{journal}{\mnras}} \textbf{\bibinfo{volume}{526}}, \bibinfo{pages}{2196--2209} (\bibinfo{year}{2023}).

\bibitem{zimmerman2024}
\bibinfo{author}{{Zimmerman}, D.~T.}, \bibinfo{author}{{Narayanan}, D.}, \bibinfo{author}{{Whitaker}, K.~E.} \& \bibinfo{author}{{Dav{\'e}}, R.}
\newblock \bibinfo{title}{{Tracing the History of Obscured Star Formation with the SIMBA Cosmological Galaxy Evolution Simulation}}.
\newblock \emph{\bibinfo{journal}{\apj}} \textbf{\bibinfo{volume}{973}}, \bibinfo{pages}{146} (\bibinfo{year}{2024}).

\bibitem{faisst2024}
\bibinfo{author}{{Faisst}, A.~L.} \& \bibinfo{author}{{Morishita}, T.}
\newblock \bibinfo{title}{{Dead or Alive? How Bursty Star Formation and Patchy Dust Can Cause Temporary Quiescence in High-redshift Galaxies}}.
\newblock \emph{\bibinfo{journal}{\apj}} \textbf{\bibinfo{volume}{971}}, \bibinfo{pages}{47} (\bibinfo{year}{2024}).

\bibitem{bowler2024}
\bibinfo{author}{{Bowler}, R.~A.~A.} \emph{et~al.}
\newblock \bibinfo{title}{{The ALMA REBELS survey: obscured star formation in massive Lyman-break galaxies at z= 4-8 revealed by the IRX-{\ensuremath{\beta}} and M$_{{\ensuremath{\star}}}$ relations}}.
\newblock \emph{\bibinfo{journal}{\mnras}} \textbf{\bibinfo{volume}{527}}, \bibinfo{pages}{5808--5828} (\bibinfo{year}{2024}).

\bibitem{wang2024}
\bibinfo{author}{{Wang}, B.} \emph{et~al.}
\newblock \bibinfo{title}{{RUBIES: Evolved Stellar Populations with Extended Formation Histories at z {\ensuremath{\sim}} 7{\textendash}8 in Candidate Massive Galaxies Identified with JWST/NIRSpec}}.
\newblock \emph{\bibinfo{journal}{\apjl}} \textbf{\bibinfo{volume}{969}}, \bibinfo{pages}{L13} (\bibinfo{year}{2024}).

\bibitem{setton2024}
\bibinfo{author}{{Setton}, D.~J.} \emph{et~al.}
\newblock \bibinfo{title}{{Little Red Dots at an Inflection Point: Ubiquitous ``V-Shaped'' Turnover Consistently Occurs at the Balmer Limit}}.
\newblock \emph{\bibinfo{journal}{arXiv e-prints}} \bibinfo{pages}{arXiv:2411.03424} (\bibinfo{year}{2024}).

\bibitem{degraaff2025_nature}
\bibinfo{author}{{de Graaff}, A.} \emph{et~al.}
\newblock \bibinfo{title}{{Efficient formation of a massive quiescent galaxy at redshift 4.9}}.
\newblock \emph{\bibinfo{journal}{Nature Astronomy}} \textbf{\bibinfo{volume}{9}}, \bibinfo{pages}{280--292} (\bibinfo{year}{2025}).

\bibitem{steinhardt2024}
\bibinfo{author}{{Steinhardt}, C.~L.} \emph{et~al.}
\newblock \bibinfo{title}{{The Highest-redshift Balmer Breaks as a Test of {\ensuremath{\Lambda}}CDM}}.
\newblock \emph{\bibinfo{journal}{\apj}} \textbf{\bibinfo{volume}{967}}, \bibinfo{pages}{172} (\bibinfo{year}{2024}).

\bibitem{weaver2023}
\bibinfo{author}{{Weaver}, J.~R.} \emph{et~al.}
\newblock \bibinfo{title}{{COSMOS2020: The galaxy stellar mass function. The assembly and star formation cessation of galaxies at 0.2< z {\ensuremath{\leq}} 7.5}}.
\newblock \emph{\bibinfo{journal}{\aap}} \textbf{\bibinfo{volume}{677}}, \bibinfo{pages}{A184} (\bibinfo{year}{2023}).

\bibitem{nc2024}
\bibinfo{author}{{Navarro-Carrera}, R.} \emph{et~al.}
\newblock \bibinfo{title}{{Burstiness in low stellar-mass Ha emitters at z\raisebox{-0.5ex}\textasciitilde2 and z\raisebox{-0.5ex}\textasciitilde4-6 from JWST medium band photometry in GOODS-S}}.
\newblock \emph{\bibinfo{journal}{arXiv e-prints}} \bibinfo{pages}{arXiv:2410.23249} (\bibinfo{year}{2024}).

\bibitem{iyer2025}
\bibinfo{author}{{Iyer}, K.~G.}, \bibinfo{author}{{Pacifici}, C.}, \bibinfo{author}{{Calistro-Rivera}, G.} \& \bibinfo{author}{{Lovell}, C.~C.}
\newblock \bibinfo{title}{{The Spectral Energy Distributions of Galaxies}}.
\newblock \emph{\bibinfo{journal}{arXiv e-prints}} \bibinfo{pages}{arXiv:2502.17680} (\bibinfo{year}{2025}).

\bibitem{Leja2019}
\bibinfo{author}{{Leja}, J.}, \bibinfo{author}{{Carnall}, A.~C.}, \bibinfo{author}{{Johnson}, B.~D.}, \bibinfo{author}{{Conroy}, C.} \& \bibinfo{author}{{Speagle}, J.~S.}
\newblock \bibinfo{title}{{How to Measure Galaxy Star Formation Histories. II. Nonparametric Models}}.
\newblock \emph{\bibinfo{journal}{\apj}} \textbf{\bibinfo{volume}{876}}, \bibinfo{pages}{3} (\bibinfo{year}{2019}).

\bibitem{carnall2019}
\bibinfo{author}{{Carnall}, A.~C.} \emph{et~al.}
\newblock \bibinfo{title}{{How to Measure Galaxy Star Formation Histories. I. Parametric Models}}.
\newblock \emph{\bibinfo{journal}{\apj}} \textbf{\bibinfo{volume}{873}}, \bibinfo{pages}{44} (\bibinfo{year}{2019}).

\bibitem{zhang2025}
\bibinfo{author}{{Zhang}, Y.}, \bibinfo{author}{{Mo}, H.~J.}, \bibinfo{author}{{Whitaker}, K.~E.} \& \bibinfo{author}{{Zhou}, S.}
\newblock \bibinfo{title}{{Testing a New Star Formation History Model from Principal Component Analysis to Facilitate Spectral Synthesis Modeling}}.
\newblock \emph{\bibinfo{journal}{\apj}} \textbf{\bibinfo{volume}{989}}, \bibinfo{pages}{179} (\bibinfo{year}{2025}).

\bibitem{speagle2014}
\bibinfo{author}{{Speagle}, J.~S.}, \bibinfo{author}{{Steinhardt}, C.~L.}, \bibinfo{author}{{Capak}, P.~L.} \& \bibinfo{author}{{Silverman}, J.~D.}
\newblock \bibinfo{title}{{A Highly Consistent Framework for the Evolution of the Star-Forming ``Main Sequence'' from z \raisebox{-0.5ex}\textasciitilde 0-6}}.
\newblock \emph{\bibinfo{journal}{\apjs}} \textbf{\bibinfo{volume}{214}}, \bibinfo{pages}{15} (\bibinfo{year}{2014}).

\bibitem{baldry2008}
\bibinfo{author}{{Baldry}, I.~K.}, \bibinfo{author}{{Glazebrook}, K.} \& \bibinfo{author}{{Driver}, S.~P.}
\newblock \bibinfo{title}{{On the galaxy stellar mass function, the mass-metallicity relation and the implied baryonic mass function}}.
\newblock \emph{\bibinfo{journal}{\mnras}} \textbf{\bibinfo{volume}{388}}, \bibinfo{pages}{945--959} (\bibinfo{year}{2008}).

\bibitem{franco2025}
\bibinfo{author}{{Franco}, M.} \emph{et~al.}
\newblock \bibinfo{title}{{Physical properties of galaxies and the UV Luminosity Function from $z\sim6$ to $z\sim14$ in COSMOS-Web}}.
\newblock \emph{\bibinfo{journal}{arXiv e-prints}} \bibinfo{pages}{arXiv:2508.04791} (\bibinfo{year}{2025}).

\bibitem{ferrara2024_z14}
\bibinfo{author}{{Ferrara}, A.}
\newblock \bibinfo{title}{{The eventful life of GS-z14-0, the most distant galaxy at redshift z = 14.32}}.
\newblock \emph{\bibinfo{journal}{\aap}} \textbf{\bibinfo{volume}{689}}, \bibinfo{pages}{A310} (\bibinfo{year}{2024}).

\bibitem{sun2016}
\bibinfo{author}{{Sun}, G.} \& \bibinfo{author}{{Furlanetto}, S.~R.}
\newblock \bibinfo{title}{{Constraints on the star formation efficiency of galaxies during the epoch of reionization}}.
\newblock \emph{\bibinfo{journal}{\mnras}} \textbf{\bibinfo{volume}{460}}, \bibinfo{pages}{417--433} (\bibinfo{year}{2016}).

\bibitem{donnan2025_efficiency}
\bibinfo{author}{{Donnan}, C.~T.}, \bibinfo{author}{{Dunlop}, J.~S.}, \bibinfo{author}{{McLure}, R.~J.}, \bibinfo{author}{{McLeod}, D.~J.} \& \bibinfo{author}{{Cullen}, F.}
\newblock \bibinfo{title}{{No evidence (yet) for increased star-formation efficiency at early times}}.
\newblock \emph{\bibinfo{journal}{\mnras}} \textbf{\bibinfo{volume}{539}}, \bibinfo{pages}{2409--2423} (\bibinfo{year}{2025}).

\bibitem{glazebrook2024}
\bibinfo{author}{{Glazebrook}, K.} \emph{et~al.}
\newblock \bibinfo{title}{{A massive galaxy that formed its stars at z {\ensuremath{\approx}} 11}}.
\newblock \emph{\bibinfo{journal}{\nat}} \textbf{\bibinfo{volume}{628}}, \bibinfo{pages}{277--281} (\bibinfo{year}{2024}).

\bibitem{carnall2024}
\bibinfo{author}{{Carnall}, A.~C.} \emph{et~al.}
\newblock \bibinfo{title}{{The JWST EXCELS survey: too much, too young, too fast? Ultra-massive quiescent galaxies at 3 < z < 5}}.
\newblock \emph{\bibinfo{journal}{\mnras}} \textbf{\bibinfo{volume}{534}}, \bibinfo{pages}{325--348} (\bibinfo{year}{2024}).

\bibitem{onoue2025}
\bibinfo{author}{{Onoue}, M.} \emph{et~al.}
\newblock \bibinfo{title}{{A post-starburst pathway for the formation of massive galaxies and black holes at z > 6}}.
\newblock \emph{\bibinfo{journal}{Nature Astronomy}}  (\bibinfo{year}{2025}).

\bibitem{sato2024}
\bibinfo{author}{{Sato}, R.~A.} \emph{et~al.}
\newblock \bibinfo{title}{{JWST/NIRSpec spectroscopy of intermediate-mass quiescent galaxies at z 3-4}}.
\newblock \emph{\bibinfo{journal}{\mnras}} \textbf{\bibinfo{volume}{534}}, \bibinfo{pages}{3552--3564} (\bibinfo{year}{2024}).

\bibitem{cutler2024}
\bibinfo{author}{{Cutler}, S.~E.} \emph{et~al.}
\newblock \bibinfo{title}{{Two Distinct Classes of Quiescent Galaxies at Cosmic Noon Revealed by JWST PRIMER and UNCOVER}}.
\newblock \emph{\bibinfo{journal}{\apjl}} \textbf{\bibinfo{volume}{967}}, \bibinfo{pages}{L23} (\bibinfo{year}{2024}).

\bibitem{pan2025}
\bibinfo{author}{{Pan}, R.} \emph{et~al.}
\newblock \bibinfo{title}{{UNCOVER/MegaScience: No Evidence of Environmental Quenching in a z$\sim$2.6 Proto-cluster}}.
\newblock \emph{\bibinfo{journal}{arXiv e-prints}} \bibinfo{pages}{arXiv:2504.06334} (\bibinfo{year}{2025}).

\bibitem{oke_gunn_1983}
\bibinfo{author}{{Oke}, J.~B.} \& \bibinfo{author}{{Gunn}, J.~E.}
\newblock \bibinfo{title}{{Secondary standard stars for absolute spectrophotometry.}}
\newblock \emph{\bibinfo{journal}{\apj}} \textbf{\bibinfo{volume}{266}}, \bibinfo{pages}{713--717} (\bibinfo{year}{1983}).

\bibitem{chabrier2003}
\bibinfo{author}{{Chabrier}, G.}
\newblock \bibinfo{title}{{Galactic Stellar and Substellar Initial Mass Function}}.
\newblock \emph{\bibinfo{journal}{\pasp}} \textbf{\bibinfo{volume}{115}}, \bibinfo{pages}{763--795} (\bibinfo{year}{2003}).

\bibitem{kroupa2002}
\bibinfo{author}{{Kroupa}, P.} \& \bibinfo{author}{{Boily}, C.~M.}
\newblock \bibinfo{title}{{On the mass function of star clusters}}.
\newblock \emph{\bibinfo{journal}{\mnras}} \textbf{\bibinfo{volume}{336}}, \bibinfo{pages}{1188--1194} (\bibinfo{year}{2002}).

\bibitem{salpeter1955}
\bibinfo{author}{{Salpeter}, E.~E.}
\newblock \bibinfo{title}{{The Luminosity Function and Stellar Evolution.}}
\newblock \emph{\bibinfo{journal}{\apj}} \textbf{\bibinfo{volume}{121}}, \bibinfo{pages}{161} (\bibinfo{year}{1955}).

\bibitem{noirot2023}
\bibinfo{author}{{Noirot}, G.} \emph{et~al.}
\newblock \bibinfo{title}{{The first large catalogue of spectroscopic redshifts in Webb's first deep field, SMACS J0723.3-7327}}.
\newblock \emph{\bibinfo{journal}{\mnras}} \textbf{\bibinfo{volume}{525}}, \bibinfo{pages}{1867--1884} (\bibinfo{year}{2023}).

\bibitem{asada2024}
\bibinfo{author}{{Asada}, Y.} \emph{et~al.}
\newblock \bibinfo{title}{{Improving photometric redshifts of Epoch of Reionization galaxies: a new transmission curve with the neutral hydrogen damped Ly$\alpha$ absorption}}.
\newblock \emph{\bibinfo{journal}{arXiv e-prints}} \bibinfo{pages}{arXiv:2410.21543} (\bibinfo{year}{2024}).

\bibitem{sarrouh2024}
\bibinfo{author}{{Sarrouh}, G. T.~E.} \emph{et~al.}
\newblock \bibinfo{title}{{Exposing Line Emission: The Systematic Differences of Measuring Galaxy Stellar Masses with JWST NIRCam Medium versus Wide Band Photometry}}.
\newblock \emph{\bibinfo{journal}{\apjl}} \textbf{\bibinfo{volume}{967}}, \bibinfo{pages}{L17} (\bibinfo{year}{2024}).

\bibitem{brammer2019}
\bibinfo{author}{{Brammer}, G.}
\newblock \bibinfo{title}{{Grizli: Grism redshift and line analysis software}}.
\newblock \bibinfo{howpublished}{Astrophysics Source Code Library, record ascl:1905.001} (\bibinfo{year}{2019}).

\bibitem{kokorev2022}
\bibinfo{author}{{Kokorev}, V.} \emph{et~al.}
\newblock \bibinfo{title}{{ALMA Lensing Cluster Survey: Hubble Space Telescope and Spitzer Photometry of 33 Lensed Fields Built with CHArGE}}.
\newblock \emph{\bibinfo{journal}{\apjs}} \textbf{\bibinfo{volume}{263}}, \bibinfo{pages}{38} (\bibinfo{year}{2022}).

\bibitem{gaiadr3}
\bibinfo{author}{{Gaia Collaboration}} \emph{et~al.}
\newblock \bibinfo{title}{Gaia data release 3 - summary of the content and survey properties}.
\newblock \emph{\bibinfo{journal}{A\&A}} \textbf{\bibinfo{volume}{674}}, \bibinfo{pages}{A1} (\bibinfo{year}{2023}).
\newblock \urlprefix\url{https://doi.org/10.1051/0004-6361/202243940}.

\bibitem{sarrouh2025}
\bibinfo{author}{{Sarrouh}, G. T.~E.} \emph{et~al.}
\newblock \bibinfo{title}{{CANUCS/Technicolor Data Release 1: Imaging, Photometry, Slit Spectroscopy, and Stellar Population Parameters}}.
\newblock \emph{\bibinfo{journal}{arXiv e-prints}} \bibinfo{pages}{arXiv:2506.21685} (\bibinfo{year}{2025}).

\bibitem{kron1980}
\bibinfo{author}{{Kron}, R.~G.}
\newblock \bibinfo{title}{{Photometry of a complete sample of faint galaxies.}}
\newblock \emph{\bibinfo{journal}{\apjs}} \textbf{\bibinfo{volume}{43}}, \bibinfo{pages}{305--325} (\bibinfo{year}{1980}).

\bibitem{brammer2008}
\bibinfo{author}{{Brammer}, G.~B.}, \bibinfo{author}{{van Dokkum}, P.~G.} \& \bibinfo{author}{{Coppi}, P.}
\newblock \bibinfo{title}{{EAZY: A Fast, Public Photometric Redshift Code}}.
\newblock \emph{\bibinfo{journal}{\apj}} \textbf{\bibinfo{volume}{686}}, \bibinfo{pages}{1503--1513} (\bibinfo{year}{2008}).

\bibitem{whitaker2011}
\bibinfo{author}{{Whitaker}, K.~E.} \emph{et~al.}
\newblock \bibinfo{title}{{The NEWFIRM Medium-band Survey: Photometric Catalogs, Redshifts, and the Bimodal Color Distribution of Galaxies out to z \raisebox{-0.5ex}\textasciitilde 3}}.
\newblock \emph{\bibinfo{journal}{\apj}} \textbf{\bibinfo{volume}{735}}, \bibinfo{pages}{86} (\bibinfo{year}{2011}).

\bibitem{fitzpatrick1999}
\bibinfo{author}{{Fitzpatrick}, E.~L.}
\newblock \bibinfo{title}{{Correcting for the Effects of Interstellar Extinction}}.
\newblock \emph{\bibinfo{journal}{\pasp}} \textbf{\bibinfo{volume}{111}}, \bibinfo{pages}{63--75} (\bibinfo{year}{1999}).

\bibitem{schlafly_finkbeiner2011}
\bibinfo{author}{{Schlafly}, E.~F.} \& \bibinfo{author}{{Finkbeiner}, D.~P.}
\newblock \bibinfo{title}{{Measuring Reddening with Sloan Digital Sky Survey Stellar Spectra and Recalibrating SFD}}.
\newblock \emph{\bibinfo{journal}{\apj}} \textbf{\bibinfo{volume}{737}}, \bibinfo{pages}{103} (\bibinfo{year}{2011}).

\bibitem{peng2010}
\bibinfo{author}{{Peng}, C.~Y.}, \bibinfo{author}{{Ho}, L.~C.}, \bibinfo{author}{{Impey}, C.~D.} \& \bibinfo{author}{{Rix}, H.-W.}
\newblock \bibinfo{title}{{Detailed Decomposition of Galaxy Images. II. Beyond Axisymmetric Models}}.
\newblock \emph{\bibinfo{journal}{\aj}} \textbf{\bibinfo{volume}{139}}, \bibinfo{pages}{2097--2129} (\bibinfo{year}{2010}).

\bibitem{akins2025}
\bibinfo{author}{{Akins}, H.~B.} \emph{et~al.}
\newblock \bibinfo{title}{{COSMOS-Web: The Overabundance and Physical Nature of ``Little Red Dots''{\textemdash}Implications for Early Galaxy and SMBH Assembly}}.
\newblock \emph{\bibinfo{journal}{\apj}} \textbf{\bibinfo{volume}{991}}, \bibinfo{pages}{37} (\bibinfo{year}{2025}).

\bibitem{larson2023}
\bibinfo{author}{{Larson}, R.~L.} \emph{et~al.}
\newblock \bibinfo{title}{{Spectral Templates Optimal for Selecting Galaxies at z > 8 with the JWST}}.
\newblock \emph{\bibinfo{journal}{\apj}} \textbf{\bibinfo{volume}{958}}, \bibinfo{pages}{141} (\bibinfo{year}{2023}).

\bibitem{boyett2024}
\bibinfo{author}{{Boyett}, K.} \emph{et~al.}
\newblock \bibinfo{title}{{Extreme emission line galaxies detected in JADES JWST/NIRSpec - I. Inferred galaxy properties}}.
\newblock \emph{\bibinfo{journal}{\mnras}} \textbf{\bibinfo{volume}{535}}, \bibinfo{pages}{1796--1828} (\bibinfo{year}{2024}).

\bibitem{papovich2025}
\bibinfo{author}{{Papovich}, C.} \emph{et~al.}
\newblock \bibinfo{title}{{Galaxies in the Epoch of Reionization Are All Bark and No Bite -- Plenty of Ionizing Photons, Low Escape Fractions}}.
\newblock \emph{\bibinfo{journal}{arXiv e-prints}} \bibinfo{pages}{arXiv:2505.08870} (\bibinfo{year}{2025}).

\bibitem{rw2005}
\bibinfo{author}{Rasmussen, C.~E.} \& \bibinfo{author}{Williams, C. K.~I.}
\newblock \emph{\bibinfo{title}{Gaussian Processes for Machine Learning}}  (\bibinfo{publisher}{The MIT Press}, \bibinfo{year}{2005}).
\newblock \urlprefix\url{https://doi.org/10.7551/mitpress/3206.001.0001}.
\newblock \eprint{https://direct.mit.edu/book-pdf/2514321/book_9780262256834.pdf}.

\bibitem{johnson2021}
\bibinfo{author}{{Johnson}, B.~D.}, \bibinfo{author}{{Leja}, J.}, \bibinfo{author}{{Conroy}, C.} \& \bibinfo{author}{{Speagle}, J.~S.}
\newblock \bibinfo{title}{{Stellar Population Inference with Prospector}}.
\newblock \emph{\bibinfo{journal}{\apjs}} \textbf{\bibinfo{volume}{254}}, \bibinfo{pages}{22} (\bibinfo{year}{2021}).

\bibitem{conroy_gunn2010}
\bibinfo{author}{{Conroy}, C.} \& \bibinfo{author}{{Gunn}, J.~E.}
\newblock \bibinfo{title}{{FSPS: Flexible Stellar Population Synthesis}}.
\newblock \bibinfo{howpublished}{Astrophysics Source Code Library, record ascl:1010.043} (\bibinfo{year}{2010}).

\bibitem{vazdekis2010}
\bibinfo{author}{{Vazdekis}, A.} \emph{et~al.}
\newblock \bibinfo{title}{{Evolutionary stellar population synthesis with MILES - I. The base models and a new line index system}}.
\newblock \emph{\bibinfo{journal}{\mnras}} \textbf{\bibinfo{volume}{404}}, \bibinfo{pages}{1639--1671} (\bibinfo{year}{2010}).

\bibitem{ferland1998}
\bibinfo{author}{{Ferland}, G.~J.} \emph{et~al.}
\newblock \bibinfo{title}{{CLOUDY 90: Numerical Simulation of Plasmas and Their Spectra}}.
\newblock \emph{\bibinfo{journal}{\pasp}} \textbf{\bibinfo{volume}{110}}, \bibinfo{pages}{761--778} (\bibinfo{year}{1998}).

\bibitem{madau1995}
\bibinfo{author}{{Madau}, P.}
\newblock \bibinfo{title}{{Radiative Transfer in a Clumpy Universe: The Colors of High-Redshift Galaxies}}.
\newblock \emph{\bibinfo{journal}{\apj}} \textbf{\bibinfo{volume}{441}}, \bibinfo{pages}{18} (\bibinfo{year}{1995}).

\bibitem{markov2025_nature}
\bibinfo{author}{{Markov}, V.} \emph{et~al.}
\newblock \bibinfo{title}{{The evolution of dust attenuation in z {\ensuremath{\approx}} 2-12 galaxies observed by JWST}}.
\newblock \emph{\bibinfo{journal}{Nature Astronomy}} \textbf{\bibinfo{volume}{9}}, \bibinfo{pages}{458--468} (\bibinfo{year}{2025}).

\bibitem{carnall2018}
\bibinfo{author}{{Carnall}, A.~C.}, \bibinfo{author}{{McLure}, R.~J.}, \bibinfo{author}{{Dunlop}, J.~S.} \& \bibinfo{author}{{Dav{\'e}}, R.}
\newblock \bibinfo{title}{{Inferring the star formation histories of massive quiescent galaxies with BAGPIPES: evidence for multiple quenching mechanisms}}.
\newblock \emph{\bibinfo{journal}{\mnras}} \textbf{\bibinfo{volume}{480}}, \bibinfo{pages}{4379--4401} (\bibinfo{year}{2018}).

\bibitem{bc03}
\bibinfo{author}{{Bruzual}, G.} \& \bibinfo{author}{{Charlot}, S.}
\newblock \bibinfo{title}{{Stellar population synthesis at the resolution of 2003}}.
\newblock \emph{\bibinfo{journal}{\mnras}} \textbf{\bibinfo{volume}{344}}, \bibinfo{pages}{1000--1028} (\bibinfo{year}{2003}).

\bibitem{bpass2009}
\bibinfo{author}{{Eldridge}, J.~J.} \& \bibinfo{author}{{Stanway}, E.~R.}
\newblock \bibinfo{title}{{Spectral population synthesis including massive binaries}}.
\newblock \emph{\bibinfo{journal}{\mnras}} \textbf{\bibinfo{volume}{400}}, \bibinfo{pages}{1019--1028} (\bibinfo{year}{2009}).

\bibitem{buchner2014}
\bibinfo{author}{{Buchner}, J.} \emph{et~al.}
\newblock \bibinfo{title}{{X-ray spectral modelling of the AGN obscuring region in the CDFS: Bayesian model selection and catalogue}}.
\newblock \emph{\bibinfo{journal}{\aap}} \textbf{\bibinfo{volume}{564}}, \bibinfo{pages}{A125} (\bibinfo{year}{2014}).

\bibitem{d'eugenio2021}
\bibinfo{author}{{D'Eugenio}, C.} \emph{et~al.}
\newblock \bibinfo{title}{{HST grism spectroscopy of z {\ensuremath{\sim}} 3 massive quiescent galaxies. Approaching the metamorphosis}}.
\newblock \emph{\bibinfo{journal}{\aap}} \textbf{\bibinfo{volume}{653}}, \bibinfo{pages}{A32} (\bibinfo{year}{2021}).

\bibitem{ma2018}
\bibinfo{author}{{Ma}, X.} \emph{et~al.}
\newblock \bibinfo{title}{{Simulating galaxies in the reionization era with FIRE-2: galaxy scaling relations, stellar mass functions, and luminosity functions}}.
\newblock \emph{\bibinfo{journal}{\mnras}} \textbf{\bibinfo{volume}{478}}, \bibinfo{pages}{1694--1715} (\bibinfo{year}{2018}).

\bibitem{pacifici2023}
\bibinfo{author}{{Pacifici}, C.} \emph{et~al.}
\newblock \bibinfo{title}{{The Art of Measuring Physical Parameters in Galaxies: A Critical Assessment of Spectral Energy Distribution Fitting Techniques}}.
\newblock \emph{\bibinfo{journal}{\apj}} \textbf{\bibinfo{volume}{944}}, \bibinfo{pages}{141} (\bibinfo{year}{2023}).

\bibitem{markov2025}
\bibinfo{author}{{Markov}, V.} \emph{et~al.}
\newblock \bibinfo{title}{{Unveiling the trends between dust attenuation and galaxy properties at $z \sim 2$-12 with JWST}}.
\newblock \emph{\bibinfo{journal}{arXiv e-prints}} \bibinfo{pages}{arXiv:2504.12378} (\bibinfo{year}{2025}).

\bibitem{suess2022}
\bibinfo{author}{{Suess}, K.~A.} \emph{et~al.}
\newblock \bibinfo{title}{{Recovering the Star Formation Histories of Recently Quenched Galaxies: The Impact of Model and Prior Choices}}.
\newblock \emph{\bibinfo{journal}{\apj}} \textbf{\bibinfo{volume}{935}}, \bibinfo{pages}{146} (\bibinfo{year}{2022}).

\bibitem{haskell2024}
\bibinfo{author}{{Haskell}, P.} \emph{et~al.}
\newblock \bibinfo{title}{{Beware the recent past: a bias in spectral energy distribution modelling due to bursty star formation}}.
\newblock \emph{\bibinfo{journal}{\mnras}} \textbf{\bibinfo{volume}{530}}, \bibinfo{pages}{L7--L12} (\bibinfo{year}{2024}).

\bibitem{kuruvanthodi2024}
\bibinfo{author}{{Kuruvanthodi}, A.} \emph{et~al.}
\newblock \bibinfo{title}{{Strong Balmer break objects at z {\ensuremath{\sim}} 7{\textendash}10 uncovered with JWST}}.
\newblock \emph{\bibinfo{journal}{\aap}} \textbf{\bibinfo{volume}{691}}, \bibinfo{pages}{A310} (\bibinfo{year}{2024}).

\bibitem{baker2025}
\bibinfo{author}{{Baker}, W.~M.} \emph{et~al.}
\newblock \bibinfo{title}{{Double Trouble: Two spectroscopically confirmed low-mass quiescent galaxies at z>5 in overdensities}}.
\newblock \emph{\bibinfo{journal}{arXiv e-prints}} \bibinfo{pages}{arXiv:2509.09761} (\bibinfo{year}{2025}).

\bibitem{planck2016}
\bibinfo{author}{{Planck Collaboration}} \emph{et~al.}
\newblock \bibinfo{title}{{Planck 2015 results. XIII. Cosmological parameters}}.
\newblock \emph{\bibinfo{journal}{\aap}} \textbf{\bibinfo{volume}{594}}, \bibinfo{pages}{A13} (\bibinfo{year}{2016}).

\bibitem{harrison_coles2012}
\bibinfo{author}{{Harrison}, I.} \& \bibinfo{author}{{Coles}, P.}
\newblock \bibinfo{title}{{Testing cosmology with extreme galaxy clusters}}.
\newblock \emph{\bibinfo{journal}{\mnras}} \textbf{\bibinfo{volume}{421}}, \bibinfo{pages}{L19--L23} (\bibinfo{year}{2012}).

\bibitem{cochrane2025}
\bibinfo{author}{{Cochrane}, R.~K.}
\newblock \bibinfo{title}{{Hierarchical assembly impedes the inference of stellar mass growth histories for individual galaxies}}.
\newblock \emph{\bibinfo{journal}{arXiv e-prints}} \bibinfo{pages}{arXiv:2509.02700} (\bibinfo{year}{2025}).

\bibitem{rg2016}
\bibinfo{author}{{Rodriguez-Gomez}, V.} \emph{et~al.}
\newblock \bibinfo{title}{{The stellar mass assembly of galaxies in the Illustris simulation: growth by mergers and the spatial distribution of accreted stars}}.
\newblock \emph{\bibinfo{journal}{\mnras}} \textbf{\bibinfo{volume}{458}}, \bibinfo{pages}{2371--2390} (\bibinfo{year}{2016}).

\bibitem{lovell2023}
\bibinfo{author}{{Lovell}, C.~C.}, \bibinfo{author}{{Harrison}, I.}, \bibinfo{author}{{Harikane}, Y.}, \bibinfo{author}{{Tacchella}, S.} \& \bibinfo{author}{{Wilkins}, S.~M.}
\newblock \bibinfo{title}{{Extreme value statistics of the halo and stellar mass distributions at high redshift: are JWST results in tension with {\ensuremath{\Lambda}}CDM?}}
\newblock \emph{\bibinfo{journal}{\mnras}} \textbf{\bibinfo{volume}{518}}, \bibinfo{pages}{2511--2520} (\bibinfo{year}{2023}).

\bibitem{marszewski2024}
\bibinfo{author}{{Marszewski}, A.}, \bibinfo{author}{{Sun}, G.}, \bibinfo{author}{{Faucher-Gigu{\`e}re}, C.-A.}, \bibinfo{author}{{Hayward}, C.~C.} \& \bibinfo{author}{{Feldmann}, R.}
\newblock \bibinfo{title}{{The High-Redshift Gas-Phase Mass{\textendash}Metallicity Relation in FIRE-2}}.
\newblock \emph{\bibinfo{journal}{\apjl}} \textbf{\bibinfo{volume}{967}}, \bibinfo{pages}{L41} (\bibinfo{year}{2024}).

\bibitem{papovich2001}
\bibinfo{author}{{Papovich}, C.}, \bibinfo{author}{{Dickinson}, M.} \& \bibinfo{author}{{Ferguson}, H.~C.}
\newblock \bibinfo{title}{{The Stellar Populations and Evolution of Lyman Break Galaxies}}.
\newblock \emph{\bibinfo{journal}{\apj}} \textbf{\bibinfo{volume}{559}}, \bibinfo{pages}{620--653} (\bibinfo{year}{2001}).

\bibitem{lorenz2025}
\bibinfo{author}{{Lorenz}, B.} \emph{et~al.}
\newblock \bibinfo{title}{{Measuring Emission Lines with JWST MegaScience Medium Bands: A New Window into Dust and Star Formation at Cosmic Noon}}.
\newblock \emph{\bibinfo{journal}{\apjl}} \textbf{\bibinfo{volume}{988}}, \bibinfo{pages}{L20} (\bibinfo{year}{2025}).

\bibitem{schaye2015}
\bibinfo{author}{{Schaye}, J.} \emph{et~al.}
\newblock \bibinfo{title}{{The EAGLE project: simulating the evolution and assembly of galaxies and their environments}}.
\newblock \emph{\bibinfo{journal}{\mnras}} \textbf{\bibinfo{volume}{446}}, \bibinfo{pages}{521--554} (\bibinfo{year}{2015}).

\bibitem{kannan2022_spectra}
\bibinfo{author}{{Kannan}, R.} \emph{et~al.}
\newblock \bibinfo{title}{{The THESAN project: predictions for multitracer line intensity mapping in the epoch of reionization}}.
\newblock \emph{\bibinfo{journal}{\mnras}} \textbf{\bibinfo{volume}{514}}, \bibinfo{pages}{3857--3878} (\bibinfo{year}{2022}).

\bibitem{pillepich2018}
\bibinfo{author}{{Pillepich}, A.} \emph{et~al.}
\newblock \bibinfo{title}{{Simulating galaxy formation with the IllustrisTNG model}}.
\newblock \emph{\bibinfo{journal}{\mnras}} \textbf{\bibinfo{volume}{473}}, \bibinfo{pages}{4077--4106} (\bibinfo{year}{2018}).

\bibitem{park2024}
\bibinfo{author}{{Park}, M.} \emph{et~al.}
\newblock \bibinfo{title}{{Widespread Rapid Quenching at Cosmic Noon Revealed by JWST Deep Spectroscopy}}.
\newblock \emph{\bibinfo{journal}{\apj}} \textbf{\bibinfo{volume}{976}}, \bibinfo{pages}{72} (\bibinfo{year}{2024}).

\bibitem{fg2018}
\bibinfo{author}{{Faucher-Gigu{\`e}re}, C.-A.}
\newblock \bibinfo{title}{{A model for the origin of bursty star formation in galaxies}}.
\newblock \emph{\bibinfo{journal}{\mnras}} \textbf{\bibinfo{volume}{473}}, \bibinfo{pages}{3717--3731} (\bibinfo{year}{2018}).

\bibitem{broussard2019}
\bibinfo{author}{{Broussard}, A.} \emph{et~al.}
\newblock \bibinfo{title}{{Star Formation Stochasticity Measured from the Distribution of Burst Indicators}}.
\newblock \emph{\bibinfo{journal}{\apj}} \textbf{\bibinfo{volume}{873}}, \bibinfo{pages}{74} (\bibinfo{year}{2019}).

\bibitem{kokorev2025}
\bibinfo{author}{{Kokorev}, V.} \emph{et~al.}
\newblock \bibinfo{title}{{CAPERS Observations of Two UV-bright Galaxies at z > 10. More Evidence for Bursting Star Formation in the Early Universe}}.
\newblock \emph{\bibinfo{journal}{\apjl}} \textbf{\bibinfo{volume}{988}}, \bibinfo{pages}{L10} (\bibinfo{year}{2025}).

\bibitem{springel2003}
\bibinfo{author}{{Springel}, V.} \& \bibinfo{author}{{Hernquist}, L.}
\newblock \bibinfo{title}{{Cosmological smoothed particle hydrodynamics simulations: a hybrid multiphase model for star formation}}.
\newblock \emph{\bibinfo{journal}{\mnras}} \textbf{\bibinfo{volume}{339}}, \bibinfo{pages}{289--311} (\bibinfo{year}{2003}).

\bibitem{witten2025_ovd}
\bibinfo{author}{{Witten}, C.} \emph{et~al.}
\newblock \bibinfo{title}{{Before its time: a remarkably evolved protocluster core at z=7.88}}.
\newblock \emph{\bibinfo{journal}{arXiv e-prints}} \bibinfo{pages}{arXiv:2507.06284} (\bibinfo{year}{2025}).

\bibitem{morishita_ovd}
\bibinfo{author}{{Morishita}, T.} \emph{et~al.}
\newblock \bibinfo{title}{{Accelerated Emergence of Evolved Galaxies in Early Overdensities at z {\ensuremath{\sim}} 5.7}}.
\newblock \emph{\bibinfo{journal}{\apj}} \textbf{\bibinfo{volume}{982}}, \bibinfo{pages}{153} (\bibinfo{year}{2025}).

\bibitem{iyer2024}
\bibinfo{author}{{Iyer}, K.~G.} \emph{et~al.}
\newblock \bibinfo{title}{{pathfinder: A Semantic Framework for Literature Review and Knowledge Discovery in Astronomy}}.
\newblock \emph{\bibinfo{journal}{\apjs}} \textbf{\bibinfo{volume}{275}}, \bibinfo{pages}{38} (\bibinfo{year}{2024}).

\end{thebibliography}

\end{document}